\documentclass[symmetry,article,accept,oneauthor,pdftex]{mdpi}

\usepackage{amsmath}
\usepackage{amsthm}
\usepackage{amssymb}
\usepackage{hyperref}
\usepackage{upgreek}

\firstpage{1}
\pubvolume{13}
\issuenum{5}
\articlenumber{773}
\pubyear{2021}
\copyrightyear{2021}
\externaleditor{Academic Editor: {Wies{\l}aw Leonski}} % For journal Automation, please change Academic Editor to "Communicated by"
\datereceived{23 March 2021}
\dateaccepted{19 April 2021}
\datepublished{29 April 2021}
\hreflink{https://\linebreak doi.org/10.3390/sym13050773} % If needed use \linebreak
\Title{Quantum Information in Neural Systems}

\TitleCitation{Quantum Information in Neural Systems}

\Author{Danko D. Georgiev 
\href{https://orcid.org/0000-0001-6846-1194}{\orcidicon}}

\AuthorNames{Danko D. Georgiev}

\AuthorCitation{Georgiev, D.D.}
\address[1]{Institute for Advanced Study, 30 Vasilaki Papadopulu Str., 9010 Varna, Bulgaria; danko@q-bits.org}

\corres{Correspondence: danko@q-bits.org}

\abstract{Identifying the physiological processes in the central nervous system that underlie our conscious experiences has been at the forefront of cognitive neuroscience. While the principles of classical physics were long found to be unaccommodating for a causally effective consciousness, the~inherent indeterminism of quantum physics, together with its characteristic dichotomy between quantum states and quantum observables, provides a fertile ground for the physical modeling of consciousness. Here, we utilize the Schr\"{o}dinger equation, together with the Planck--Einstein relation between energy and frequency, in order to determine the appropriate quantum dynamical timescale of conscious processes. Furthermore, with the help of a simple two-qubit toy model we illustrate the importance of non-zero interaction Hamiltonian for the generation of quantum entanglement and manifestation of observable correlations between different measurement outcomes. Employing a quantitative measure of entanglement based on Schmidt decomposition, we show that quantum evolution governed only by internal Hamiltonians for the individual quantum subsystems preserves quantum coherence of separable initial quantum states, but eliminates the possibility of any interaction and quantum entanglement. The~presence of non-zero interaction Hamiltonian, however, allows for decoherence of the individual quantum subsystems along with their mutual interaction and quantum entanglement. The~presented results show that quantum coherence of individual subsystems cannot be used for cognitive binding because it is a physical mechanism that leads to separability and non-interaction. In contrast, quantum interactions with their associated decoherence of individual subsystems are instrumental for dynamical changes in the quantum entanglement of the composite quantum state vector and manifested correlations of different observable outcomes. Thus, fast decoherence timescales could assist cognitive binding through quantum entanglement across extensive neural networks in the brain cortex.}


\begin{document}
%%%%%%%%%%%%%%%%%%%%%%%%%%%%%%%%%%%%%%%%%%
\section{\label{sec:intro}Introduction}

We exist in the universe and, consequently, obey its physical laws
whatever those physical laws may be \cite{Duhem1954}. The~principles
of classical physics \cite{Susskind2013,Landau1960,Kibble2004,Morin2008,Strauch2009}, however, describe a deterministic clockwork
world that is unable to accommodate causally effective conscious experiences
\cite{Georgiev2013,Georgiev2017,Georgiev2019c}. This leads to insurmountable
problems with the theory of \mbox{evolution \cite{Darwin2006,Dawkins2004,Hodge2009}}, including a lack of explanation of
how consciousness could be tolerated by natural selection~\cite{James1879,James1890,Popper1983}.
Fortunately, the~discovery of quantum mechanics in 1920s provided
a radically different picture of the physical world in which the fabric
of reality is comprised of quantum probability amplitudes for potential
physical events \cite{Feynman1948,Feynman2014,Feynman2013}, whose actual occurrence
is decided indeterministically by the inherent propensity of quantum
systems to produce a definite physical outcome upon measurement \cite{Dirac1967,vonNeumann1955,Susskind2014,Mermin1985}.
This creates a dichotomy between what exists in the form of quantum
states, and what can be observed in the form of quantum observables.
Since consciousness exists but is not observable, reductive identification
of consciousness with quantum information contained in quantum brain
states explains the inner privacy of conscious experiences \cite{Georgiev2020a}
and justifies the use of bra-ket notation for writing mental states \cite{Georgiev2020e}.
For example, there are dozens of papers on Schr\"{o}dinger's cat, Wigner's friend or Everett's many worlds/many minds interpretation of quantum mechanics that write $\vert\raisebox{-1pt}{$\ddot\smile$}\rangle $ or $\vert\raisebox{-1pt}{$\ddot\frown$}\rangle $ to indicate mental states of happiness or sadness, or to indicate the occurrence or nonoccurrence of certain conscious experiences~\cite{Tegmark2000,Everett1957,Albert1992,Lockwood1996,Hemmo2003,Wigner1962,Squires1990,Barrett1995,Whitaker2000,Fuchs2013,Frauchiger2018,Pusey2018}.
The~reductive identification of consciousness with quantum information implies that the mental
states have to satisfy the axioms of vectors in Hilbert space and obey the temporal dynamics prescribed by the Schr\"{o}dinger
equation with a physically admissible Hamiltonian. This approach allows
one to obtain mathematically precise answers to questions about the
nature of consciousness using the tools of quantum information theory.

In this work, we address a number of important questions related to
the quantum dynamics of consciousness, which has to be governed by
the Schr\"{o}dinger equation.
The~level of difficulty and accessibility of the theoretical discourse is carefully adjusted for the needs of a broad audience comprised of interdisciplinary researchers whose main expertise is in biomedical sciences but are keen to learn the basics of quantum information theory.
The~general outline of the presentation
is as follows---in Section~\ref{sec:axioms}, we introduce concisely
the basic postulates of quantum mechanics. Next, in \mbox{Section~\ref{sec:model}},
we construct a simple two-qubit toy model whose quantum dynamics illustrates
the various quantum concepts defined in Hilbert space. Then, in Section~\ref{sec:timescale}, we utilize the Planck--Einstein relation, which~connects energy and frequency, in order to characterize the appropriate
quantum timescale for conscious processes involving interaction energies
that exceed the energy of the thermal noise. Next, we elaborate on
the concepts of quantum entanglement in Section~\ref{sec:entanglement}
and quantum coherence in Section~\ref{sec:coherence}, with an emphasis
on how these concepts relate to the process of decoherence and the
presence or absence of quantum interactions. In \mbox{Sections~\ref{sec:observables}
and \ref{sec:disentangle}}, we derive the main results with regard
to the importance of internal Hamiltonians or interaction Hamiltonians
for the temporal evolution of initially separable quantum states into
quantum entangled states. In Section~\ref{sec:condensation}, we examine
the differential effects of quantum coherence and quantum entanglement
on cognitive binding. Finally, we wrap up the presentation by discussing
how the various quantum information theoretic notions relate to conscious
processes that extend across neural networks in the brain cortex.

\section{\label{sec:axioms}Basic Postulates of Quantum Mechanics}

Applying quantum mechanics to study any problem in natural sciences
requires a minimal familiarity with the fundamental quantum mechanical
postulates \cite{Dirac1967,vonNeumann1955,Susskind2014}. In~order
to make the present exposition self-contained, we will concisely summarize
those postulates, known as Dirac--von Neumann axioms \cite{Bhaumik2020},
before we use them to construct and analyze a simple two-qubit toy
model.
\begin{Axiom}
(State) The quantum physical state of a closed system is completely
described by a unit state vector $|\Psi\rangle$ in complex-valued
Hilbert space $\mathcal{H}$.
\end{Axiom}
\begin{Axiom}
(Composition) The Hilbert space of a composite quantum system comprised
from $k$~components is a tensor product of the state spaces of the
component subsystems $\mathcal{H}=\mathcal{H}_{1}\otimes\mathcal{H}_{2}\otimes\ldots\otimes\mathcal{H}_{k}$.
\end{Axiom}
\begin{Axiom}
(Observables) To every observable physical property $A$ there exists
an associated Hermitian operator $\hat{A}=\hat{A}^{\dagger}$, which
acts on the Hilbert space of states $\mathcal{H}$. The~eigenvalues
$\lambda_{A}$ of the operator $\hat{A}$ are the possible values
of the observable physical property.
\end{Axiom}
\begin{Axiom}
\textls[-15]{(Born rule) The expectation value $\langle\hat{A}\rangle$ of a measured
quantum observable $\hat{A}$ is given by the inner product with the
current quantum state $\vert\Psi\rangle$ of the physical system,
namely $\langle\hat{A}\rangle=\langle\Psi\vert\hat{A}\vert\Psi\rangle$.}
\end{Axiom}
\begin{Axiom}
(Dynamics) The time evolution of a closed physical system obeys the Schr\"{o}dinger~equation
\begin{equation}
\imath\hbar\frac{\partial}{\partial t}|\Psi\rangle=\hat{H}|\Psi\rangle ,
\end{equation}
where the Hamiltonian $\hat{H}=\hat{H}^{\dagger}$ is the observable
corresponding to the total energy of the system.
\end{Axiom}

With the use of the matrix exponential function \cite{Bellman1987},
the general solution of the Schr\"{o}dinger equation could be written
in the form
\begin{equation}
|\Psi(t)\rangle=e^{-\frac{\imath}{\hbar}\hat{H}t}|\Psi(0)\rangle ,
\end{equation}
where $|\Psi(0)\rangle$ is the initial quantum state at time $t=0$.
Thus, the~Hamiltonian $\hat{H}$ is the generator of time translation that evolves
the initial quantum state forward in time. Exactly because of this
great importance of the Hamiltonian for the ensuing quantum dynamics,
throughout this work we will be interested to delineate the differences
in action of \emph{internal Hamiltonians}, which operate only on the
reduced Hilbert subspaces of component subsystems, and \emph{interaction
Hamiltonians}, which operate on the tensor product Hilbert space of
the composite system.

The Schr\"{o}dinger equation, which comprises the core of quantum theory
\cite{Schrodinger1926,Schrodinger1928,Berezin1991}, is~linear and obeys the superposition
principle \cite{Georgiev2017}. This means that given any two solutions
$|\Psi_{1}\rangle$ and $|\Psi_{2}\rangle$ of the Schr\"{o}dinger equation,
we can construct an infinite number of solutions, which are linear
combinations of the given two, namely, $|\Psi_{s}\rangle=\alpha_{1}|\Psi_{1}\rangle+\alpha_{2}|\Psi_{2}\rangle$.
Thus,~the~linearity of the Schr\"{o}dinger equation could be viewed as
the underlying reason why the quantum states of physical systems form
a Hilbert space. Furthermore, the~Schr\"{o}dinger equation is unitary,
which is an essential ingredient in the proofs of important quantum
no-go theorems (such as the no-cloning theorem \cite{Wootters1982})
that characterize the distinctive properties of quantum information.
Therefore, there is a precise mathematical sense in which a physical
theory of consciousness should be considered \emph{quantum} only if
its basic tenets originate from the Schr\"{o}dinger equation \cite{Georgiev2013,Georgiev2017},
but not from putative violations of the Schr\"{o}dinger equation.

\section{\label{sec:model}Minimal Quantum Toy Model}

\begin{figure}[t!]
\includegraphics[width=140mm]{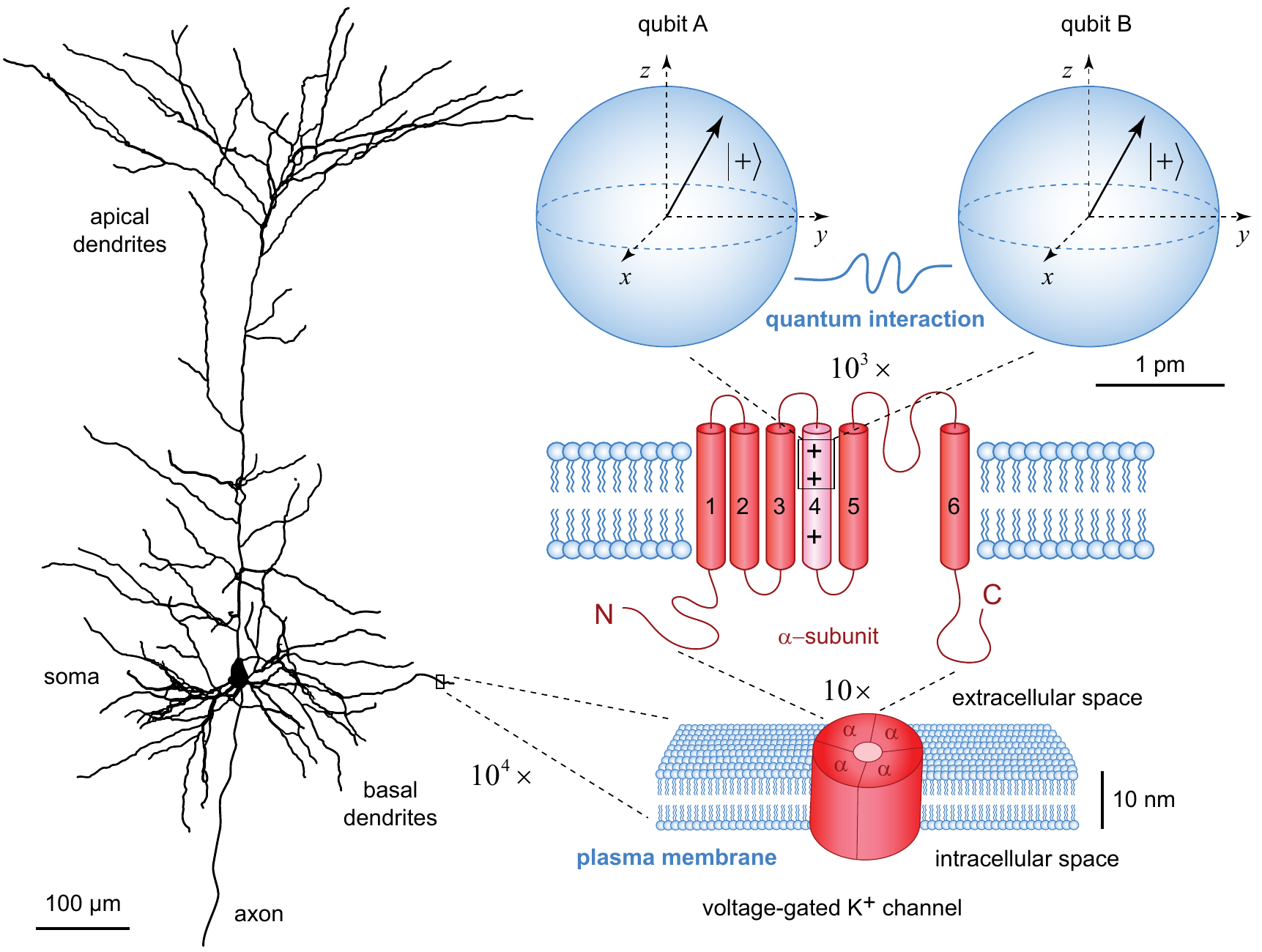}
\caption{\label{fig:1}Different levels of organization of physical processes
within the central nervous system. At the microscopic scale, the~brain cortex is composed
of neurons, which form neural networks. The~morphology of the rendered pyramidal neuron (NMO\_09565) from layer~5 of rat motor cortex (\url{http://NeuroMorpho.Org}; accessed on 19 April 2021) reflects the functional specialization of cable-like neuronal projections (dendrites and axon).
At the nanoscale, the~electric activity of neurons is generated by voltage-gated
ion channels, which are inserted in the neuronal plasma membrane.
As an example of ion channel is shown a single voltage-gated K\protect\textsuperscript{+}~channel composed of four protein \mbox{$\alpha$-subunits}.
Each subunit has six $\alpha$-helices traversing the plasma membrane.
The 4th $\alpha$-helix is positively charged and acts as voltage sensor.
At the picoscale, individual elementary electric charges within the
protein voltage sensor could be modeled as qubits represented by Bloch
spheres. For the diameter of each qubit is used the Compton wavelength of electron.
Consecutive magnifications from micrometer ($\upmu$m) to picometer (pm) scale are indicated by $\times$ symbol.}
\end{figure}

To substantiate the abstract quantum concepts in Hilbert space \cite{Akhiezer1993,Birman1986},
we will construct a toy model whose quantum dynamics could be solved
analytically and plotted for visual inspection. Because we will study
quantum interactions, we need at least two quantum systems. Furthermore,
because we would like the model to be biologically relevant (Figure~\ref{fig:1}), we will consider the quantum dynamics of electrons \cite{Dirac1928a,Dirac1928b,Thaller1992},
which are elementary particles with spin-$\frac{1}{2}$ thereby acting
as minimal quantum bits of information, \mbox{or qubits \cite{Hayashi2015}}.
The simplest possible composite quantum system that is capable of
capturing the process of entanglement and decoherence is one composed
of two non-identical interacting qubits in a uniform static magnetic
field $\vec{B}$ aligned in the $z$-direction.

\subsection{Hamiltonian of the Toy Model}

Let $\mathcal{H}_{A}$ be the two-dimensional complex Hilbert space
of the first qubit $A$ and $\mathcal{H}_{B}$ be the two-dimensional
complex Hilbert space of the second qubit $B$. Then according to
the Dirac--von Neumann axioms of quantum mechanics \cite{Dirac1967,vonNeumann1955},
the composite two-qubit quantum system resides in a four-dimensional
complex Hilbert space $\mathcal{H}_{AB}=\mathcal{H}_{A}\otimes\mathcal{H}_{B}$,
where~$\otimes$ denotes the tensor product (also known as Kronecker
product of matrices). The~eigenvectors of the observable $\hat{S}_{z,i}$
describing the spin component along the $z$-axis for the individual
Hilbert spaces can be written as $\left\{ |\uparrow_{z}\rangle_{i},|\downarrow_{z}\rangle_{i}\right\} $
where $i=A,B$. The~corresponding eigenvalues appear as scaling factors in the relations
\begin{align}
\hat{S}_{z,i}| & \uparrow_{z}\rangle_{i}=\frac{\hbar}{2}|\uparrow_{z}\rangle_{i} \\
\hat{S}_{z,i}| & \downarrow_{z}\rangle_{i}=-\frac{\hbar}{2}|\downarrow_{z}\rangle_{i} ,
\end{align}
where the reduced Planck constant is $\hbar=1.055\times10^{-34}$
J s. With the use of the Pauli spin matrices, $\hat{\sigma}_{x}$,
$\hat{\sigma}_{y}$ and $\hat{\sigma}_{z}$, we can define all three
different spin components as $\hat{S}_{x,i}=\frac{\hbar}{2}\hat{\sigma}_{x}$,
$\hat{S}_{y,i}=\frac{\hbar}{2}\hat{\sigma}_{y}$ and $\hat{S}_{z,i}=\frac{\hbar}{2}\hat{\sigma}_{z}$ in the respective Hilbert spaces $\mathcal{H}_{A}$ and $\mathcal{H}_{B}$.

\textls[-25]{In the composite four-dimensional complex Hilbert space $\mathcal{H}_{AB}$,
we can explicitly write quantum states or quantum observables in matrix
form using a chosen basis. For the description of the two-qubit toy model,
we will adopt the spin $zz$ basis, $\left\{ |\uparrow_{z}\uparrow_{z}\rangle,|\uparrow_{z}\downarrow_{z}\rangle,|\downarrow_{z}\uparrow_{z}\rangle,|\downarrow_{z}\downarrow_{z}\rangle\right\} $,}
where for economy of notation we have used ordering from left to right
to indicate which subsystem $i=A,B$ is referred to, namely
\begin{align}
|\uparrow_{z}\uparrow_{z}\rangle & \equiv|\uparrow_{z}\rangle_{A}|\uparrow_{z}\rangle_{B} \\
|\uparrow_{z}\downarrow_{z}\rangle & \equiv|\uparrow_{z}\rangle_{A}|\downarrow_{z}\rangle_{B} \\
|\downarrow_{z}\uparrow_{z}\rangle & \equiv|\downarrow_{z}\rangle_{A}|\uparrow_{z}\rangle_{B} \\
|\downarrow_{z}\downarrow_{z}\rangle & \equiv|\downarrow_{z}\rangle_{A}|\downarrow_{z}\rangle_{B} .
\end{align}

Then, the~individual spin observables in $\mathcal{H}_{AB}$ are given
by product operators
\begin{equation}
\hat{S}_{x,A}=\frac{\hbar}{2}\hat{\sigma}_{x}\otimes\hat{I}_{B}=\frac{\hbar}{2}\left(\begin{array}{cc}
0 & 1\\
1 & 0
\end{array}\right)\otimes\left(\begin{array}{cc}
1 & 0\\
0 & 1
\end{array}\right)=\frac{\hbar}{2}\left(\begin{array}{cccc}
0 & 0 & 1 & 0\\
0 & 0 & 0 & 1\\
1 & 0 & 0 & 0\\
0 & 1 & 0 & 0
\end{array}\right)
\end{equation}
\begin{equation}
\hat{S}_{x,B}=\frac{\hbar}{2}\hat{I}_{A}\otimes\hat{\sigma}_{x}=\frac{\hbar}{2}\left(\begin{array}{cc}
1 & 0\\
0 & 1
\end{array}\right)\otimes\left(\begin{array}{cc}
0 & 1\\
1 & 0
\end{array}\right)=\frac{\hbar}{2}\left(\begin{array}{cccc}
0 & 1 & 0 & 0\\
1 & 0 & 0 & 0\\
0 & 0 & 0 & 1\\
0 & 0 & 1 & 0
\end{array}\right)
\end{equation}
\begin{equation}
\hat{S}_{y,A}=\frac{\hbar}{2}\hat{\sigma}_{y}\otimes\hat{I}_{B}=\frac{\hbar}{2}\left(\begin{array}{cc}
0 & -\imath\\
\imath & 0
\end{array}\right)\otimes\left(\begin{array}{cc}
1 & 0\\
0 & 1
\end{array}\right)=\frac{\hbar}{2}\left(\begin{array}{cccc}
0 & 0 & -\imath & 0\\
0 & 0 & 0 & -\imath\\
\imath & 0 & 0 & 0\\
0 & \imath & 0 & 0
\end{array}\right)
\end{equation}
\begin{equation}
\hat{S}_{y,B}=\frac{\hbar}{2}\hat{I}_{A}\otimes\hat{\sigma}_{y}=\frac{\hbar}{2}\left(\begin{array}{cc}
1 & 0\\
0 & 1
\end{array}\right)\otimes\left(\begin{array}{cc}
0 & -\imath\\
\imath & 0
\end{array}\right)=\frac{\hbar}{2}\left(\begin{array}{cccc}
0 & -\imath & 0 & 0\\
\imath & 0 & 0 & 0\\
0 & 0 & 0 & -\imath\\
0 & 0 & \imath & 0
\end{array}\right)
\end{equation}
\begin{equation}
\hat{S}_{z,A}=\frac{\hbar}{2}\hat{\sigma}_{z}\otimes\hat{I}_{B}=\frac{\hbar}{2}\left(\begin{array}{cc}
1 & 0\\
0 & -1
\end{array}\right)\otimes\left(\begin{array}{cc}
1 & 0\\
0 & 1
\end{array}\right)=\frac{\hbar}{2}\left(\begin{array}{cccc}
1 & 0 & 0 & 0\\
0 & 1 & 0 & 0\\
0 & 0 & -1 & 0\\
0 & 0 & 0 & -1
\end{array}\right)
\end{equation}
\begin{equation}
\hat{S}_{z,B}=\frac{\hbar}{2}\hat{I}_{A}\otimes\hat{\sigma}_{z}=\frac{\hbar}{2}\left(\begin{array}{cc}
1 & 0\\
0 & 1
\end{array}\right)\otimes\left(\begin{array}{cc}
1 & 0\\
0 & -1
\end{array}\right)=\frac{\hbar}{2}\left(\begin{array}{cccc}
1 & 0 & 0 & 0\\
0 & -1 & 0 & 0\\
0 & 0 & 1 & 0\\
0 & 0 & 0 & -1
\end{array}\right) .
\end{equation}
The magnetic moments of the two qubits are $\vec{\mu}_{1}=\gamma_{1}\vec{S}_{1}$
and $\vec{\mu}_{2}=\gamma_{2}\vec{S}_{2}$, where the gyromagnetic
ratios are $\gamma_{1}=g_{1}\frac{q_{1}}{2m_{1}}$ and $\gamma_{2}=g_{2}\frac{q_{2}}{2m_{2}}$,
with $g$-factors $g_{1}$ and $g_{2}$, electric~charges $q_{1}$
and $q_{2}$, and masses $m_{1}$ and $m_{2}$.

For a qubit that is realized by a spinning electron, we have
\begin{equation}
\vec{\mu}=\gamma_{e}\vec{S}=g_{e}\frac{q_{e}}{2m_{e}}\vec{S}=g_{e}\frac{\mu_{B}}{\hbar}\vec{S},
\end{equation}
where the electron $g$-factor is $g_{e}=-2.00231930436256$ and the Bohr magneton is \linebreak
\mbox{$\mu_{B}=\frac{q_{e}\hbar}{2m_{e}}\approx9.27\times10^{-24}$}~J/Tesla.

If the internal Hamiltonians for each of the two qubits $A$ and $B$
are respectively
\begin{align}
\hat{H}_{A} & =-\mu_{1}\cdot\vec{B}\\
\hat{H}_{B} & =-\mu_{2}\cdot\vec{B}
\end{align}
and the interaction between the two qubits is given by
\begin{equation}
\hat{H}_{\textrm{int}}=\xi\left(\vec{S}_{1}\cdot\vec{S}_{2}\right)
\end{equation}
we can write the total Hamiltonian as
\begin{equation}
\hat{H}=\hat{H}_{A}+\hat{H}_{B}+\hat{H}_{\textrm{int}}.
\end{equation}
After substitution and expanding the inner products, we have
\begin{align}
\hat{H} & =-\mu_{1}\cdot\vec{B}-\mu_{2}\cdot\vec{B}+\xi\left(\vec{S}_{1}\cdot\vec{S}_{2}\right)\nonumber \\
& =\Omega_{1}\vec{S}_{z,A}+\Omega_{2}\vec{S}_{z,B}+\xi\left(\hat{S}_{x,A}\hat{S}_{x,B}+\hat{S}_{y,A}\hat{S}_{y,B}+\hat{S}_{z,A}\hat{S}_{z,B}\right)\nonumber \\
& =\frac{\hbar}{2}\Omega_{1}\left(\begin{array}{cccc}
1 & 0 & 0 & 0\\
0 & 1 & 0 & 0\\
0 & 0 & -1 & 0\\
0 & 0 & 0 & -1
\end{array}\right)+\frac{\hbar}{2}\Omega_{2}\left(\begin{array}{cccc}
1 & 0 & 0 & 0\\
0 & -1 & 0 & 0\\
0 & 0 & 1 & 0\\
0 & 0 & 0 & -1
\end{array}\right)+\frac{\hbar^{2}}{4}\xi\left(\begin{array}{cccc}
1 & 0 & 0 & 0\\
0 & -1 & 2 & 0\\
0 & 2 & -1 & 0\\
0 & 0 & 0 & 1
\end{array}\right)\nonumber \\
& =\left(\begin{array}{cccc}
E_{+}+E_{s} & 0 & 0 & 0\\
0 & E_{-}-E_{s} & 2E_{s} & 0\\
0 & 2E_{s} & -E_{-}-E_{s} & 0\\
0 & 0 & 0 & -E_{+}+E_{s}
\end{array}\right) , \label{eq:Hamiltonian}
\end{align}
where $E_{+}=\frac{1}{2}\hbar\left(\Omega_{1}+\Omega_{2}\right)$,
$E_{-}=\frac{1}{2}\hbar\left(\Omega_{1}-\Omega_{2}\right)$ and $E_{s}=\frac{\hbar^{2}}{4}\xi$.
The corresponding angular frequencies are $\omega_{+}=\frac{1}{2}\left(\Omega_{1}+\Omega_{2}\right)$,
$\omega_{-}=\frac{1}{2}\left(\Omega_{1}-\Omega_{2}\right)$ and $\omega_{s}=\frac{\hbar}{4}\xi$.

\subsection{Energy Eigenstates and Eigenvalues of the Toy Model}

The eigenvalues and eigenvectors of the Hamiltonian correspond to
physical energies that enter in the general solution of the Schr\"{o}dinger
equation \cite{Susskind2014}. The~energy eigenvalues of the concrete
toy Hamiltonian \eqref{eq:Hamiltonian} are found to be
\begin{align}
E_{1} & =E_{+}+E_{s}\label{eq:E1}\\
E_{2} & =-E_{s}-\sqrt{E_{-}^{2}+4E_{s}^{2}}\label{eq:E2}\\
E_{3} & =-E_{s}+\sqrt{E_{-}^{2}+4E_{s}^{2}}\label{eq:E3}\\
E_{4} & =-E_{+}+E_{s}. \label{eq:E4}
\end{align}
Their corresponding energy eigenvectors expressed in the spin $zz$ basis are given by
\begin{align}
|E_{1}\rangle & =|\uparrow_{z}\uparrow_{z}\rangle\\
|E_{2}\rangle & =-C_{1}|\uparrow_{z}\downarrow_{z}\rangle+C_{2}\vert\downarrow_{z}\uparrow_{z}\rangle\\
|E_{3}\rangle & =C_{2}|\uparrow_{z}\downarrow_{z}\rangle+C_{1}|\downarrow_{z}\uparrow_{z}\rangle\\
|E_{4}\rangle & =|\downarrow_{z}\downarrow_{z}\rangle ,
\end{align}
where for economy of notation we have set
\begin{align}
C_{1} & =\sin\left[\frac{1}{2}\arccos\left(\frac{E_{-}}{\sqrt{E_{-}^{2}+4E_{s}^{2}}}\right)\right]=\frac{1}{\sqrt{2}}\sqrt{1-\frac{E_{-}}{\sqrt{E_{-}^{2}+4E_{s}^{2}}}}\\
C_{2} & =\cos\left[\frac{1}{2}\arccos\left(\frac{E_{-}}{\sqrt{E_{-}^{2}+4E_{s}^{2}}}\right)\right]=\frac{1}{\sqrt{2}}\sqrt{1+\frac{E_{-}}{\sqrt{E_{-}^{2}+4E_{s}^{2}}}}.
\end{align}
The mathematical relationship between eigenvectors and their eigenvalues
allows for easy computation of the action of the Hamiltonian operator,
namely
\begin{equation}
\hat{H}|E_{n}\rangle=E_{n}|E_{n} \rangle .
\end{equation}
Hence, if we express the initial composite two-qubit quantum state
$|\Psi\rangle$ in the energy basis
\begin{equation}
|\Psi\rangle=\hat{I}|\Psi\rangle=\sum_{n}\mathcal{\hat{P}}\left(E_{n}\right)\vert\Psi\rangle=\sum_{n}|E_{n}\rangle\langle E_{n}|\Psi\rangle=\sum_{n}\alpha_{n}|E_{n} \rangle ,
\end{equation}
we can propagate it in time using the Schr\"{o}dinger equation
\begin{equation}
\imath\hbar\frac{\partial}{\partial t}|\Psi\rangle=\hat{H}|\Psi\rangle=\hat{H}\sum_{n}\alpha_{n}|E_{n}\rangle=\sum_{n}E_{n}\alpha_{n}|E_{n}\rangle .
\end{equation}
\textls[-25]{Explicitly rewriting the quantum state in the energy basis gives a
system of differential~equations}
\begin{equation}
\imath\hbar\frac{\partial}{\partial t}\left(\begin{array}{c}
\alpha_{1}\\
\alpha_{2}\\
\vdots\\
\alpha_{n}
\end{array}\right)=\left(\begin{array}{c}
E_{1}\alpha_{1}\\
E_{2}\alpha_{2}\\
\vdots\\
E_{n}\alpha_{n}
\end{array}\right) .
\end{equation}
Before we solve these differential equations, we can divide both sides
by $\imath\hbar$ and work only with the angular frequencies
\begin{equation}
\frac{\partial}{\partial t}\left(\begin{array}{c}
\alpha_{1}\\
\alpha_{2}\\
\vdots\\
\alpha_{n}
\end{array}\right)=-\imath\left(\begin{array}{c}
\omega_{1}\alpha_{1}\\
\omega_{2}\alpha_{2}\\
\vdots\\
\omega_{n}\alpha_{n}
\end{array}\right) , \label{eq:Schr-omega}
\end{equation}
where $\omega_{1}=E_{1}/\hbar$, $\omega_{2}=E_{2}/\hbar$, $\ldots$,
$\omega_{n}=E_{n}/\hbar$. Quantum physicists usually take the approach
of working in units in which $\hbar=1$, hence energy and angular frequency
become equivalent by the Planck--Einstein relation $E=\hbar\omega$.
However, our chosen approach is more general as it does not require
fixing the physical units. Also, compared to energy, the~angular frequencies
are more appropriate for characterizing the dynamic timescale of the
composite quantum system.

\subsection{Quantum Dynamics of the State Vector}

The solutions of the Schr\"{o}dinger Equation \eqref{eq:Schr-omega} are
easily found in the form
\begin{equation}
\alpha_{n}(t)=\alpha_{n}(0)e^{-\imath\omega_{n}t},
\end{equation}
with the immediate interpretation that $\alpha_{n}(0)$ is the initial
quantum probability amplitude of the state $|E_{n}\rangle$ at $t=0$.

Then, the~general solution of the time-dependent Schr\"{o}dinger equation
becomes
\begin{align}
|\Psi(t)\rangle & =\sum_{n}\alpha_{n}(0)e^{-\imath\omega_{n}t}|E_{n}\rangle\\
& =\alpha_{1}(0)e^{-\imath\omega_{1}t}|E_{1}\rangle+\alpha_{2}(0)e^{-\imath\omega_{2}t}|E_{2}\rangle+\alpha_{3}(0)e^{-\imath\omega_{3}t}|E_{3}\rangle+\alpha_{4}(0)e^{-\imath\omega_{4}t}|E_{4} \rangle ,
\end{align}
where from Equations \eqref{eq:E1}--\eqref{eq:E4} we obtain
\begin{align}
\omega_{1} & =\omega_{+}+\omega_{s}\label{eq:w1}\\
\omega_{2} & =-\omega_{s}-\sqrt{\omega_{-}^{2}+4\omega_{s}^{2}}\label{eq:w2}\\
\omega_{3} & =-\omega_{s}+\sqrt{\omega_{-}^{2}+4\omega_{s}^{2}}\label{eq:w3}\\
\omega_{4} & =-\omega_{+}+\omega_{s}.\label{eq:w4}
\end{align}
The coefficients of the energy eigenstates also can be fully re-written
in terms of the angular~frequencies
\begin{align}
C_{1} & =\frac{1}{\sqrt{2}}\sqrt{1-\frac{\omega_{-}}{\sqrt{\omega_{-}^{2}+4\omega_{s}^{2}}}}\\
C_{2} & =\frac{1}{\sqrt{2}}\sqrt{1+\frac{\omega_{-}}{\sqrt{\omega_{-}^{2}+4\omega_{s}^{2}}}}.
\end{align}
Thus, the~general quantum state $|\Psi(t)\rangle$ at any time $t$ is expressed
as
\begin{align}
|\Psi(t)\rangle & =\alpha_{1}(0)e^{-\imath\omega_{1}t}|\uparrow_{z}\uparrow_{z}\rangle-\left(C_{1}\alpha_{2}(0)e^{-\imath\omega_{2}t}-C_{2}\alpha_{3}(0)e^{-\imath\omega_{3}t}\right)|\uparrow_{z}\downarrow_{z}\rangle\nonumber \\
& +\left(C_{2}\alpha_{2}(0)e^{-\imath\omega_{2}t}+C_{1}\alpha_{3}(0)e^{-\imath\omega_{3}t}\right)|\downarrow_{z}\uparrow_{z}\rangle+\alpha_{4}(0)e^{-\imath\omega_{4}t}|\downarrow_{z}\downarrow_{z}\rangle . \label{eq:sol-zz}
\end{align}
The initial energy quantum probability amplitudes can be further re-written in their
exponential form as
\begin{equation}
\alpha_{i}(0)=\left|\alpha_{i}(0)\right|e^{\imath\varphi_{i}},
\end{equation}
where the principal arguments are
\begin{equation}
\varphi_{i}\equiv\textrm{Arg}\left[\alpha_{i}(0) \right] .
\end{equation}

The quantum state $|\Psi(t)\rangle$ describes what exists in the quantum world. The
quantum state, however, is not an observable entity \cite{Busch1997}.
In order to determine what can be measured, we need to consider the expectation
values of quantum observables.

\section{\label{sec:timescale}Quantum Dynamic Timescale}

To determine the relevant dynamic timescale, we need to account for
Landauer's principle, according to which there is a minimum possible
amount of energy $E_{\min}$ required to erase one bit of information
\cite{Landauer1961} or to transmit one bit successfully across a
noisy quantum channel affected by thermal noise \cite{Levitin1998,Georgiev2020f}
\begin{equation}
E_{\min}=k_{B}T\ln2\approx2.97\times10^{-21}\quad\textrm{J} ,
\end{equation}
where $k_{B}=1.38\times10^{-23}$ J/K is the Boltzmann constant and
$T=310$ K is the physiological body temperature of humans. From the Planck--Einstein
relation $E=\hbar\omega$, we can determine the minimal angular frequency
\begin{equation}
\omega_{\min}=\frac{k_{B}T\ln2}{\hbar}\approx2.8\times10^{13}\quad\textrm{rad}/\textrm{s} .
\end{equation}
This means that there will be $\frac{\omega_{\min}}{2\pi}\times10^{-12}\approx4.5$
full rotations for time period of 1 ps. The~time period for 1 full
rotation is $\frac{2\pi}{\omega_{\min}}\approx0.22$ ps.

As an alternative calculation, consider the interaction of the Bohr magneton with an external
magnetic field of 3~Tesla, which is routinely used for functional
magnetic resonance imaging (fMRI) of the human brain, to obtain an interaction
energy
\begin{equation}
\frac{1}{2}g_{e}\mu_{B}B\approx2.78\times10^{-23}\quad\textrm{J} ,
\end{equation}
that is about 100 times smaller than the Landauer's limit.
The corresponding angular frequency is
\begin{equation}
\Omega_{1}=\Omega_{2}=\frac{g_{e}\mu_{B}B}{2\hbar}\approx2.64\times10^{11}\quad\textrm{rad}/\textrm{s} .
\end{equation}
This means that there will be $\frac{\Omega_{1}}{2\pi}\times10^{-10}\approx4.2$
full rotations for time period of 100 ps. The~time period for 1 full
rotation is $\frac{2\pi}{\omega_{\textrm{int}}}\approx23.8$ ps.
One possible interpretation of the latter result is that the MRI experimental technique is by a factor of $\approx$100 slower that the predicted  quantum characteristic time of neural processes related to consciousness.

The main conclusion is that for typical energies that are comparable
to the thermal energy, the~rotation of the state vector in Hilbert
space occurs with time period at a subpicosecond timescale. For energies
that are an order of magnitude lower than the thermal energy, the
dynamic timescale is picoseconds, for energies that are two orders
of magnitude lower than the thermal energy, the~dynamic timescale
is tens of picoseconds, and~so on. Therefore, the~Planck--Einstein
relation $E=\hbar\omega$ that appears in the Schr\"{o}dinger equation
fixes the dynamic timescale of quantum processes to picosecond timescale
or faster, which~is in the realm of quantum chemistry \cite{Georgiev2017,Georgiev2020e}.
For complex biomolecules such as proteins, which catalyze life-sustaining physiological activities, key functional role is played by hydrogen bonds whose energy is an order of magnitude lower compared to covalent bonds. Consequently, the~quantum transport of energy in protein $\alpha$-helices  due to conformational bending of hydrogen bonded peptide groups occurs at picosecond \mbox{timescale \cite{Georgiev2019a,Georgiev2019b,Georgiev2020c}}, as opposed to femtosecond electron transport due to destruction or creation of covalent \mbox{bonds \cite{Marcus1956,Vos1993,Beck1992,Beck1996,Beck1998}.}
Quantum theories of consciousness that require operation at longer
timescales, such as milliseconds, are incompatible with the essentials
established by the Planck--Einstein relation and the Schr\"{o}dinger equation.

Here, we would like to emphasize that there are contrived methods
based on quantum beats that could realize destructive quantum interference
in order to slow down the quantum dynamics of the expectation value of
some quantum observables, however,~those~methods are not evolutionary
plausible as they reduce the computational power of the brain. To appreciate the meaning of this criticism, consider
the operation of computer processors realized on silicon chips: a
processor that operates at MHz (microsecond timescale) can perform
only $10^{6}$ operations per second, whereas a processor that operates
at GHz (nanosecond timescale) can perform $10^{9}$ operations per
second. The~faster the processor is, the~more computational power
it has. That is why modern computing technologies aim toward processors
operating at THz (picosecond timescale), where quantum effects become
dominant, rather than the slower timescale of milliseconds or even seconds,
where~the computation can be performed manually e.g., by sliding beads
on abacus.

\section{\label{sec:entanglement}Quantum Entanglement}

Quantum entanglement entails interdependence between different component
subsystems, which form a composite quantum system \cite{Schrodinger1935}.
Although the quantum entanglement could be manifested in the form
of correlations between observable outcomes, the~converse is not true,
namely the lack of correlations between observable outcomes does not
imply the lack of entanglement. This is because quantum entanglement
is defined for the quantum state, which is a vector in Hilbert space,
whereas the probabilities for occurrence of definite measurement outcomes
pertain to quantum observables, which are operators on the Hilbert
space.

Incompatible (non-commuting) observables cannot be measured simultaneously
with a single experimental setting of the measurement apparatus. This
means that in the act of quantum measurement only a set of compatible
(commuting) quantum observables can be determined to have a definite
value. Those quantum observables that are not measured do not have
a definite value and consequently there are no actual measurement
outcomes from which could be extracted correlations. If one is careful
to distinguish between what is actual and what is counterfactual,
quantum mechanics allows for exact prediction of the possible correlations
that would have been observed had the necessary quantum measurements
been performed.
\begin{Definition}
(Separable state) A bipartite composite quantum state $|\Psi\rangle\in\mathcal{H}_{A}\otimes\mathcal{H}_{B}$
is called separable if and only if it can be written as a tensor product
\cite{Gudder2020b}
\begin{equation}
|\Psi\rangle=|\psi\rangle_{A}\otimes|\psi\rangle_{B}.
\end{equation}

\end{Definition}

\begin{Definition}
(Entangled state) A bipartite composite quantum state $|\Psi\rangle\in\mathcal{H}_{A}\otimes\mathcal{H}_{B}$
is called entangled if and only if it cannot be written as a tensor
product \cite{Gudder2020b}
\begin{equation}
|\Psi\rangle\neq|\psi\rangle_{A}\otimes|\psi\rangle_{B}.
\end{equation}

\end{Definition}

\begin{Theorem}
(Schmidt decomposition) Let \textup{$\left\{ |i\rangle_{A}\right\} $}
be a basis for $\mathcal{H}_{A}$ and \textup{$\left\{ |j\rangle_{B}\right\} $}
be a basis for $\mathcal{H}_{B}$. Then, every bipartite composite
quantum state $|\Psi\rangle\in\mathcal{H}_{A}\otimes\mathcal{H}_{B}$
can be expressed in $\left\{ |i\rangle_{A}|j\rangle_{B}\right\} $
basis as follows
\begin{equation}
|\Psi\rangle=\sum_{i}\sum_{j}c_{ij}|i\rangle_{A}|j\rangle_{B}.
\end{equation}
Next, construct the matrix $\hat{C}=\left(c_{ij}\right)$ and perform singular
value decomposition in the form
\begin{equation}
\hat{C}=\hat{U}\hat{\Lambda}\hat{V}^{\dagger},
\end{equation}
where $\hat{U}$ and $\hat{V}^{\dagger}$ are unitary matrices, and
$\hat{\Lambda}$ is a diagonal matrix with non-negative singular values
sorted in descending order $\lambda_{1}\geq\lambda_{2}\geq\ldots\geq\lambda_{s}\geq0$
also referred to as Schmidt coefficients. Thus, the~bipartite composite
quantum state becomes
\begin{equation}
|\Psi\rangle=\sum_{s}\lambda_{s}|\psi\rangle_{A}\otimes|\psi\rangle_{B},
\end{equation}
where $|\psi\rangle_{A}=\hat{U}|i\rangle_{A}$ and $|\psi\rangle_{B}=\hat{V}^{\dagger}|j\rangle_{B}$
is the Schmidt basis \cite{Miszczak2011}. The~state is entangled
if its Schmidt rank is greater than 1. Otherwise, the~state is separable.
\end{Theorem}

\begin{Theorem}
The singular values of a Hermitian matrix $\hat{A}=\hat{A}^{\dagger}$
are the absolute values of the eigenvalues of $\hat{A}$.\end{Theorem}
\begin{proof}
Every Hermitian matrix has a complete set of eigenvectors, and all of its eigenvalues are real.
This allows spectral decomposition
\begin{equation}
\hat{A}=\hat{U}\hat{\Lambda}\hat{U}^{\dagger},
\end{equation}
where $\hat{U}$ is unitary matrix and $\hat{\Lambda}=\hat{\Lambda}^{\dagger}$
is a diagonal matrix with real entries. Decompose the diagonal matrix
as $\hat{\Lambda}=\left|\hat{\Lambda}\right|\textrm{sign}\hat{\Lambda}$.
The matrix $\hat{V}^{\dagger}=\textrm{sign}\hat{\Lambda}\hat{U}^{\dagger}$
is unitary because $\hat{U}^{\dagger}$ is unitary. Therefore, the
singular value decomposition of $\hat{A}$ is
\begin{equation}
\hat{A}=\hat{U}\left|\hat{\Lambda}\right|\hat{V}^{\dagger}.
\end{equation}
\end{proof}
\begin{Definition}
The entanglement number $e\left(\Psi\right)$ of a quantum state $|\Psi\rangle$
is determined by the Schmidt coefficients using the formula \cite{Gudder2020a}
\begin{equation}
e\left(\Psi\right)=\sqrt{1-\sum_{s}\lambda_{s}^{4}}.\label{eq:Gudder}
\end{equation}
\end{Definition}

\begin{Theorem}
The Hermitian matrix $\hat{C}\hat{C}^{\dagger}$ is positive semidefinite, namely
$\langle x|\hat{C}\hat{C}^{\dagger}|x\rangle\geq0$ for all vectors~$x$.
The eigenvalues of the Hermitian matrix $\hat{C}\hat{C}^{\dagger}=\hat{U}\hat{\Lambda}\hat{V}^{\dagger}\hat{V}\hat{\Lambda}\hat{U}^{\dagger}=\hat{U}\hat{\Lambda}^{2}\hat{U}^{\dagger}$
are equal to~$\lambda_{s}^{2}$. Therefore, an efficient way to compute
the entanglement number without recourse to singular value decomposition
is to use \cite{Gudder2020a}
\begin{equation}
e\left(\Psi\right)=\sqrt{1-\textrm{Tr}\left(\hat{C}\hat{C}^{\dagger}\hat{C}\hat{C}^{\dagger}\right)}.\label{eq:entanglement}
\end{equation}
The entanglement number could be normalized in the interval $[0,1]$
as follows
\begin{equation}
\tilde{e}\left(\Psi\right)=\frac{e\left(\Psi\right)}{e\left(\Psi\right)_{\max}},
\end{equation}
where $e\left(\Psi\right)_{\max}$ is the maximal possible value,
which is determined by the dimension $n$ of the Hilbert space of
the smallest subsystem, namely $n=\min\left[d\left(\mathcal{H}_{A}\right),d\left(\mathcal{H}_{B}\right)\right]$.
In the maximally entangled state, the~Schmidt decomposition has $n$
eigenvalues equal to $\frac{1}{\sqrt{n}}$, resulting in
\begin{equation}
e\left(\Psi\right)_{\max}=\sqrt{1-n\frac{1}{n^{2}}}=\sqrt{\frac{n-1}{n}}.
\end{equation}

\end{Theorem}
The normalized quantum entanglement number $e\left(\Psi\right)$ in
the toy model could be computed from the eigenvalues of $\hat{C}\hat{C}^{\dagger}$ to be
\begin{align}
\tilde{e}\left(\Psi\right) & =2\Big\{\left|\alpha_{1}^{2}(0)\alpha_{4}^{2}(0)\right|+\left|\alpha_{2}^{2}(0)\alpha_{3}^{2}(0)\right|+C_{1}^{2}C_{2}^{2}\left(\left|\alpha_{2}^{4}(0)\right|-4\left|\alpha_{2}^{2}(0)\alpha_{3}^{2}(0)\right|+\left|\alpha_{3}^{4}(0)\right|\right)\nonumber \\
& \quad-2C_{+}C_{-}\left|\alpha_{1}(0)\alpha_{2}(0)\alpha_{3}(0)\alpha_{4}(0)\right|\cos\left[\left(\omega_{1}-\omega_{2}-\omega_{3}+\omega_{4}\right)t-\varphi_{1}+\varphi_{2}+\varphi_{3}-\varphi_{4}\right]\nonumber \\
& \quad+2C_{1}C_{2}\left|\alpha_{1}(0)\alpha_{2}^{2}(0)\alpha_{4}(0)\right|\cos\left[\left(\omega_{1}-2\omega_{2}+\omega_{4}\right)t-\varphi_{1}+2\varphi_{2}-\varphi_{4}\right]\nonumber \\
& \quad-2C_{1}C_{2}\left|\alpha_{1}(0)\alpha_{3}^{2}(0)\alpha_{4}(0)\right|\cos\left[\left(\omega_{1}-2\omega_{3}+\omega_{4}\right)t-\varphi_{1}+2\varphi_{3}-\varphi_{4}\right]\nonumber \\
& \quad-2C_{1}C_{2}C_{+}C_{-}\left(\left|\alpha_{2}^{3}(0)\alpha_{3}(0)\right|-\left|\alpha_{2}(0)\alpha_{3}^{3}(0)\right|\right)\cos\left[\left(\omega_{2}-\omega_{3}\right)t-\varphi_{2}+\varphi_{3}\right]\nonumber \\
& \quad-2C_{1}^{2}C_{2}^{2}\left|\alpha_{2}^{2}(0)\alpha_{3}^{2}(0)\right|\cos\left[2\left(\omega_{2}-\omega_{3}\right)t-2\left(\varphi_{2}-\varphi_{3}\right)\right]\Big\}^{\frac{1}{2}}.
\end{align}

\section{\label{sec:coherence}Quantum Coherence}

Quantum coherence and decoherence are frequently mentioned in discussions
on the feasibility of quantum approaches to consciousness \cite{Tegmark2000,Tegmark2015}.
Because the visibility of quantum interference patterns requires quantum
superpositions \cite{Qureshi2019a,Qureshi2019b}, a careless wording
may say that quantum coherence is indicative of present quantum superpositions,
whereas~decoherence is indicative of absent quantum superpositions.
Unfortunately, such statements cannot be literally correct and would
appear to be based on misunderstanding the vector nature of quantum
states. Mathematically, a vector can always be decomposed into a superposition
of other vectors, which means that the concept of quantum superposition
is not an absolute property, but a basis-dependent one \cite{Georgiev2017}.
Therefore, it is meaningless to talk about the presence or absence
of quantum superpositions without explicitly stating the basis in
which those superpositions are considered.
\begin{Definition}
(Quantum coherence) The $\ell_{1}$ norm of coherence $\mathcal{C}$
is a basis-dependent quantitative measure of quantum coherence of
an $n\times n$ dimensional density matrix $\hat{\rho}$ defined with
the use of the sum of all off-diagonal moduli \cite{Bera2015}
\begin{equation}
\mathcal{C}=\frac{1}{n-1}\sum_{i\neq j}\left|\rho_{ij}\right|.
\end{equation}
The quantum coherence is bounded in the interval $0\leq\mathcal{C}\leq1$.\end{Definition}
\begin{Example}
Consider a qubit in the state $\vert\uparrow_{z}\rangle$. The~density
matrix of the qubit in spin $z$ basis is
\begin{equation}
\hat{\rho}=\vert\uparrow_{z}\rangle\langle\uparrow_{z}\vert=\left(\begin{array}{cc}
1 & 0\end{array}\right)\left(\begin{array}{c}
1\\
0
\end{array}\right)=\left(\begin{array}{cc}
1 & 0\\
0 & 0
\end{array}\right) .
\end{equation}
All off-diagonal elements of the density matrix $\hat{\rho}$ expressed
in spin $z$ basis are zero, hence $\mathcal{C}=0$, indicating that
the state is not superposed in that basis. However, the~same density
matrix re-written in spin $x$ basis becomes
\begin{equation}
\hat{\rho}=\frac{1}{\sqrt{2}}\left(\vert\uparrow_{x}\rangle+\vert\downarrow_{x}\rangle\right)\frac{1}{\sqrt{2}}\left(\langle\uparrow_{x}\vert+\langle\downarrow_{x}\vert\right)=\frac{1}{2}\left(\begin{array}{cc}
1 & 1\end{array}\right)\left(\begin{array}{c}
1\\
1
\end{array}\right)=\frac{1}{2}\left(\begin{array}{cc}
1 & 1\\
1 & 1
\end{array}\right) .
\end{equation}
All off-diagonal elements of the density matrix $\hat{\rho}$ expressed
in spin $x$ basis are $\frac{1}{2}$, hence $\mathcal{C}=1$, indicating
that the state is maximally superposed in that basis.
\end{Example}
It should be emphasized that the disappearance of quantum coherence
merely by change of basis of a pure state \cite{Streltsov2018} is
not the main physical phenomenon studied by decoherence theory. Instead,
decoherence refers to the loss of quantum purity and conversion of
pure quantum states into mixed quantum states \cite{Finkelstein1993,Zeh1997,Zeh2000,Zurek2003}.
In this latter process, the~coherent quantum superpositions, which
can manifest visible quantum interference patterns under suitable
choice of measurement basis, become converted into incoherent quantum
superpositions, which cannot manifest visible quantum interference
patterns. In general, the~loss of visibility of interference patterns
is not abrupt, but occurs gradually with the loss of quantum purity.
\begin{Definition}
(Quantum purity) The quantum purity $\gamma$ of an $n\times n$ dimensional
density matrix $\hat{\rho}$ is defined as
\begin{equation}
\gamma=\textrm{Tr}\left(\hat{\rho}^{2}\right).
\end{equation}
The quantum purity is bounded in the interval $\frac{1}{n}\leq\gamma\leq1$.\end{Definition}
\begin{Example}
A maximally mixed state with minimal purity $\gamma=\frac{1}{2}$
is realized by a qubit with density matrix
\begin{equation}
\hat{\rho}=\frac{1}{2}\left(\vert\uparrow_{z}\rangle\langle\uparrow_{z}\vert+\vert\downarrow_{z}\rangle\langle\downarrow_{z}\vert\right)=\frac{1}{2}\left(\vert\uparrow_{x}\rangle\langle\uparrow_{x}\vert+\vert\downarrow_{x}\rangle\langle\downarrow_{x}\vert\right)=\frac{1}{2}\left(\begin{array}{cc}
1 & 0\\
0 & 1
\end{array}\right) .
\end{equation}
The maximally mixed state is incoherent in every basis, meaning that
the $\ell_{1}$ norm of coherence $\mathcal{C}$ is zero in every
basis. As a result, quantum measurement in any basis will find the
spin with equal probability being in either of the two possible directions,
up or down. Furthermore, quantum superpositions may still be present,
even though the maximally mixed state will not manifest visible quantum
interference patterns in any basis. For that latter reason, such quantum
superpositions are called {incoherent}.
\end{Example}
To understand where the incoherent quantum superpositions reside,
we have to consider the quantum mechanical concept of purification.
\begin{Definition}
(Purification) Every mixed state given by a density matrix $\hat{\rho}=\sum_{i}p_{i}\vert i\rangle\langle i\vert$
acting on finite-dimensional Hilbert space $\mathcal{H}_{A}$ with
basis $\left\{ \vert i\rangle\right\} $ can be viewed as the reduced
state of some pure state $|\Psi\rangle\in\mathcal{H}_{A}\otimes\mathcal{H}_{B}$,
where $\mathcal{H}_{B}$ is another copy of $\mathcal{H}_{A}$ with
basis $\left\{ \vert i'\rangle\right\} $.\end{Definition}
\begin{proof}
Every density matrix $\hat{\rho}\in\mathcal{H}_{A}$ can be spectrally
decomposed in terms of its eigenvalues $p_{i}$ and corresponding
set of orthonormal eigenvectors $\left\{ \vert i\rangle\right\} $.
Working with the basis $\left\{ \vert i\rangle\right\} $ massively
simplifies the calculations because the density matrix $\hat{\rho}$
becomes diagonal in that basis. Then, consider the state $|\Psi\rangle=\sum_{i}\sqrt{p_{i}}|i\rangle\otimes\vert i'\rangle$.
Explicit calculation of the partial trace gives
\begin{align}
\hat{\rho}
& =\textrm{Tr}_{B}\vert\Psi\rangle\langle\Psi\vert \nonumber\\
& =\textrm{Tr}_{B}\left(\sum_{i}\sqrt{p_{i}}|i\rangle\otimes\vert i'\rangle\sum_{j}\sqrt{p_{j}}\langle j\vert\otimes\langle j'\vert\right)\nonumber \\
& =\textrm{Tr}_{B}\sum_{i}\sum_{j}\sqrt{p_{i}p_{j}}|i\rangle\langle j\vert\otimes\vert i'\rangle\langle j'\vert \nonumber \\
& =\sum_{i}\sum_{j}\delta_{ij}\sqrt{p_{i}p_{j}}|i\rangle\langle j\vert \nonumber \\
& =\sum_{i}p_{i}\vert i\rangle\langle i\vert.
\end{align}
\end{proof}
\begin{Example}
Consider two qubits $A$ and $B$ comprising the pure maximally entangled
(anti-correlated) quantum state
\begin{equation}
|\Psi\rangle_{AB}=\frac{1}{\sqrt{2}}\left(\vert\uparrow_{z}\downarrow_{z}\rangle+\vert\downarrow_{z}\uparrow_{z}\rangle\right)\label{eq:ent-1}.
\end{equation}
The composite density matrix is pure
\begin{equation}
\hat{\rho}_{AB}=\hat{\rho}_{AB}^{2}=\frac{1}{2}\left(\begin{array}{cccc}
0 & 0 & 0 & 0\\
0 & 1 & 1 & 0\\
0 & 1 & 1 & 0\\
0 & 0 & 0 & 0
\end{array}\right) ,
\end{equation}
whereas the reduced density matrices of each qubit are maximally mixed
\begin{equation}
\hat{\rho}_{A}=\textrm{Tr}_{B}\left(\hat{\rho}_{AB}\right)=\hat{\rho}_{B}=\textrm{Tr}_{A}\left(\hat{\rho}_{AB}\right)=\frac{1}{2}\left(\begin{array}{cc}
1 & 0\\
0 & 1
\end{array}\right) .
\end{equation}
Thus, the~quantum superposition is seen as non-zero off-diagonal elements
only in the composite density matrix $\hat{\rho}_{AB}$, but not in
the reduced density matrices $\hat{\rho}_{A}$ and $\hat{\rho}_{B}$.
\end{Example}

\begin{Example}
Consider two qubits $A$ and $B$ comprising the pure maximally entangled
(correlated) quantum state
\begin{equation}
|\Psi\rangle_{AB}=\frac{1}{\sqrt{2}}\left(\vert\uparrow_{z}\uparrow_{z}\rangle+\vert\downarrow_{z}\downarrow_{z}\rangle\right)\label{eq:ent-2}.
\end{equation}
Again, the~composite density matrix is pure
\begin{equation}
\hat{\rho}_{AB}=\hat{\rho}_{AB}^{2}=\frac{1}{2}\left(\begin{array}{cccc}
1 & 0 & 0 & 1\\
0 & 0 & 0 & 0\\
0 & 0 & 0 & 0\\
1 & 0 & 0 & 1
\end{array}\right) ,
\end{equation}
whereas the reduced density matrices of each qubit are maximally mixed
\vspace{12pt}
\begin{equation}
\hat{\rho}_{A}=\textrm{Tr}_{B}\left(\hat{\rho}_{AB}\right)=\hat{\rho}_{B}=\textrm{Tr}_{A}\left(\hat{\rho}_{AB}\right)=\frac{1}{2}\left(\begin{array}{cc}
1 & 0\\
0 & 1
\end{array}\right) .
\end{equation}
The reduced density matrices $\hat{\rho}_{A}$ and $\hat{\rho}_{B}$
are exactly the same as in the case when the composite state is given
by \eqref{eq:ent-1}. Comparison of the the composite states \eqref{eq:ent-1}
and \eqref{eq:ent-2} shows that the measurement outcomes of the $z$-components
of the two spins will be correlated or anti-correlated, respectively.
Therefore, knowing only the reduced density matrices of the components
does not provide a complete description of the composite system, whereas
knowing the composite state vector does.
\end{Example}
In the context of decoherence theory, \emph{quantum coherence} is
used with the intended meaning of \emph{maximal purity}. Keeping this
clarification in mind, it could be said that the composite two-qubit
system is quantum coherent because its purity is maximal, $\gamma=1$,
whereas each of the component qubits is incoherent because its purity
is minimal, $\gamma=\frac{1}{2}$. Thus, quantum entanglement leads
to decoherence and the two processes go hand by hand. Conversely,
the two component qubits cannot be individually maximally coherent
(pure) and quantum entangled at the same time. Instead, if each of
the two qubits is in a pure state, then the composite state is separable.
\begin{Example}
Consider each of the two qubits $A$ and $B$ being in coherent quantum
superposition $\frac{1}{\sqrt{2}}\left(\vert\uparrow_{z}\rangle+\vert\downarrow_{z}\rangle\right)$.
The composite density matrix is pure
\begin{equation}
\hat{\rho}_{AB}=\hat{\rho}_{AB}^{2}=\frac{1}{4}\left(\begin{array}{cccc}
1 & 1 & 1 & 1\\
1 & 1 & 1 & 1\\
1 & 1 & 1 & 1\\
1 & 1 & 1 & 1
\end{array}\right)
\end{equation}
and the reduced density matrices of each qubit are also pure
\begin{equation}
\hat{\rho}_{A}=\hat{\rho}_{A}^{2}=\textrm{Tr}_{B}\left(\hat{\rho}_{AB}\right)=\hat{\rho}_{B}=\hat{\rho}_{B}^{2}=\textrm{Tr}_{A}\left(\hat{\rho}_{AB}\right)=\frac{1}{2}\left(\begin{array}{cc}
1 & 1\\
1 & 1
\end{array}\right) .
\end{equation}
The separability of the composite state follows from the purity of
the components
\begin{equation}
|\Psi\rangle_{AB}=\frac{1}{\sqrt{2}}\left(\vert\uparrow_{z}\rangle+\vert\downarrow_{z}\rangle\right)\otimes\frac{1}{\sqrt{2}}\left(\vert\uparrow_{z}\rangle+\vert\downarrow_{z}\rangle\right)=\vert\uparrow_{x}\rangle\otimes\vert\uparrow_{x}\rangle .
\end{equation}

\end{Example}
From the preceding examples, it should be clear that quantum mechanics
contains two very different kinds of relationships between composite
systems and component systems. For separable states, both the composite
system and the component systems have pure quantum states. This implies
that it is possible to write state vectors for both the composite
system and the component systems. For quantum entangled states, it
is only the composite system that has a pure quantum state, whereas
the component systems are necessarily described by mixed quantum states.
Because mixed quantum states can only be represented in the form of
a density matrix, but not a quantum state vector, this implies that
the components of quantum entangled states cannot be completely described
in isolation. Indeed, the~reduced density matrices that describe the
components of quantum entangled states can be used only for computing
the quantum probabilities of outcomes from local measurements performed
upon the given component. However, the~reduced density matrices are
useless for computing existing correlations between different components
\cite{Georgiev2017}.

\section{\label{sec:observables}Measurement of Quantum Observables}

So far, we have discussed the importance of the state vector $|\Psi(t)\rangle$
of the composite system and have solved the Schr\"{o}dinger equation for
the two-qubit toy model. Although, the~state vector $|\Psi(t)\rangle$
is not an observable \cite{Busch1997}, it allows us to determine
the expectation values of any physical observable $\hat{A}$ that
could be measured at any point in time $t$. The~important thing to
keep in mind is that non-commuting quantum observables are incompatible
with each other and cannot be measured at the same time. This means
that without knowing which observable is actually measured, the~quantum
dynamics of the expectation value of any quantum observable $\langle\hat{A}\rangle=\langle\Psi(t)\vert\hat{A}|\Psi(t)\rangle$
is to be considered as counterfactual, namely, it describes probability
distributions of physical events that could have happened provided
that the quantum observable $\hat{A}$ were measured. Thus, quantum
observables do not necessarily reflect what exists, but only what
could be observed \cite{Georgiev2020a,Georgiev2020e}.

\subsection{Quantum Observables in Spin $zz$ Basis}

The probabilities for measuring the composite system in each of the
spin $zz$ basis states from the set $\left\{ |\uparrow_{z}\uparrow_{z}\rangle,|\uparrow_{z}\downarrow_{z}\rangle,|\downarrow_{z}\uparrow_{z}\rangle,|\downarrow_{z}\downarrow_{z}\rangle\right\} $
are given by the expectation values of the corresponding projection
operators according to the Born rule \cite{Born1955}, namely
\begin{align}
\textrm{Prob}\left(\uparrow_{z}\uparrow_{z}\right) & =\langle\Psi(t)|\hat{\mathcal{P}}\left(\uparrow_{z}\uparrow_{z}\right)|\Psi(t)\rangle=\langle\Psi(t)|\uparrow_{z}\uparrow_{z}\rangle\langle\uparrow_{z}\uparrow_{z}|\Psi(t)\rangle=\left|\langle\uparrow_{z}\uparrow_{z}|\Psi(t)\rangle\right|^{2}\\
\textrm{Prob}\left(\uparrow_{z}\downarrow_{z}\right) & =\langle\Psi(t)|\hat{\mathcal{P}}\left(\uparrow_{z}\downarrow_{z}\right)|\Psi(t)\rangle=\langle\Psi(t)|\uparrow_{z}\downarrow_{z}\rangle\langle\uparrow_{z}\downarrow_{z}|\Psi(t)\rangle=\left|\langle\uparrow_{z}\downarrow_{z}|\Psi(t)\rangle\right|^{2}\\
\textrm{Prob}\left(\downarrow_{z}\uparrow_{z}\right) & =\langle\Psi(t)|\hat{\mathcal{P}}\left(\downarrow_{z}\uparrow_{z}\right)|\Psi(t)\rangle=\langle\Psi(t)|\downarrow_{z}\uparrow_{z}\rangle\langle\downarrow_{z}\uparrow_{z}|\Psi(t)\rangle=\left|\langle\downarrow_{z}\uparrow_{z}|\Psi(t)\rangle\right|^{2}\\
\textrm{Prob}\left(\downarrow_{z}\downarrow_{z}\right) & =\langle\Psi(t)|\hat{\mathcal{P}}\left(\downarrow_{z}\downarrow_{z}\right)|\Psi(t)\rangle=\langle\Psi(t)|\downarrow_{z}\downarrow_{z}\rangle\langle\downarrow_{z}\downarrow_{z}|\Psi(t)\rangle=\left|\langle\downarrow_{z}\downarrow_{z}|\Psi(t)\rangle\right|^{2}.
\end{align}
Explicit calculation for the observable quantum probabilities
gives
\begin{align}
\textrm{Prob}\left(\uparrow_{z}\uparrow_{z}\right) & =\left|\alpha_{1}(0)\right|^{2}\label{eq:zz-1}\\
\textrm{Prob}\left(\uparrow_{z}\downarrow_{z}\right) & =C_{1}^{2}\left|\alpha_{2}(0)\right|^{2}-2C_{1}C_{2}\left|\alpha_{2}(0)\alpha_{3}(0)\right|\cos\left[\left(\omega_{2}-\omega_{3}\right)t-\varphi_{2}+\varphi_{3}\right]+C_{2}^{2}\left|\alpha_{3}(0)\right|^{2}\label{eq:zz-2}\\
\textrm{Prob}\left(\downarrow_{z}\uparrow_{z}\right) & =C_{2}^{2}\left|\alpha_{2}(0)\right|^{2}+2C_{1}C_{2}\left|\alpha_{2}(0)\alpha_{3}(0)\right|\cos\left[\left(\omega_{2}-\omega_{3}\right)t-\varphi_{2}+\varphi_{3}\right]+C_{1}^{2}\left|\alpha_{3}(0)\right|^{2}\label{eq:zz-3}\\
\textrm{Prob}\left(\downarrow_{z}\downarrow_{z}\right) & =\left|\alpha_{4}(0)\right|^{2}.\label{eq:zz-4}
\end{align}
It is easy to see that the four quantum probabilities sum up to unity
since $C_{1}^{2}+C_{2}^{2}=1$ and $\sum_{n}\left|\alpha_{n}(0)\right|^{2}=1$.

\textls[-25]{In order to perform computer simulations, one can plug in directly
different quantum probability amplitudes $\alpha_{n}(0)$ for the
initial superposition of energy states \mbox{$|\Psi(0)\rangle=\sum_{n}\alpha_{n}(0)|E_{n}\rangle$}
and then} observe the quantum dynamics that follows. In the case when
the initial quantum state $|\Psi(0)\rangle$ is given in a different
basis, such as an eigenvector of the evolving quantum observable,
then one first needs to determine the initial quantum probability
amplitudes $\alpha_{n}(0)$ for the energy states using inner products
\begin{equation}
|\Psi(0)\rangle=\hat{I}|\Psi(0)\rangle=\sum_{n}|E_{n}\rangle\langle E_{n}|\Psi(0)\rangle=\sum_{n}\alpha_{n}(0)|E_{n}\rangle
\end{equation}
and then proceed with the computer simulation. As an explicit example,
we present the initial energy quantum probability amplitudes for each
of the spin $zz$ basis states in the energy basis
\begin{align}
|\uparrow_{z}\uparrow_{z}\rangle & =|E_{1}\rangle\\
|\uparrow_{z}\downarrow_{z}\rangle & =-C_{1}|E_{2}\rangle+C_{2}|E_{3}\rangle\\
|\downarrow_{z}\uparrow_{z}\rangle & =C_{2}|E_{2}\rangle+C_{1}|E_{3}\rangle\\
|\downarrow_{z}\downarrow_{z}\rangle & =|E_{4}\rangle .
\end{align}
Once we have the initial quantum probability amplitudes in the energy
basis, we can plot the quantum dynamics of the expectation values
of the quantum observables given in Equations~\eqref{eq:zz-1}--\eqref{eq:zz-4}.

\begin{figure}[t!]
\includegraphics[width=140mm]{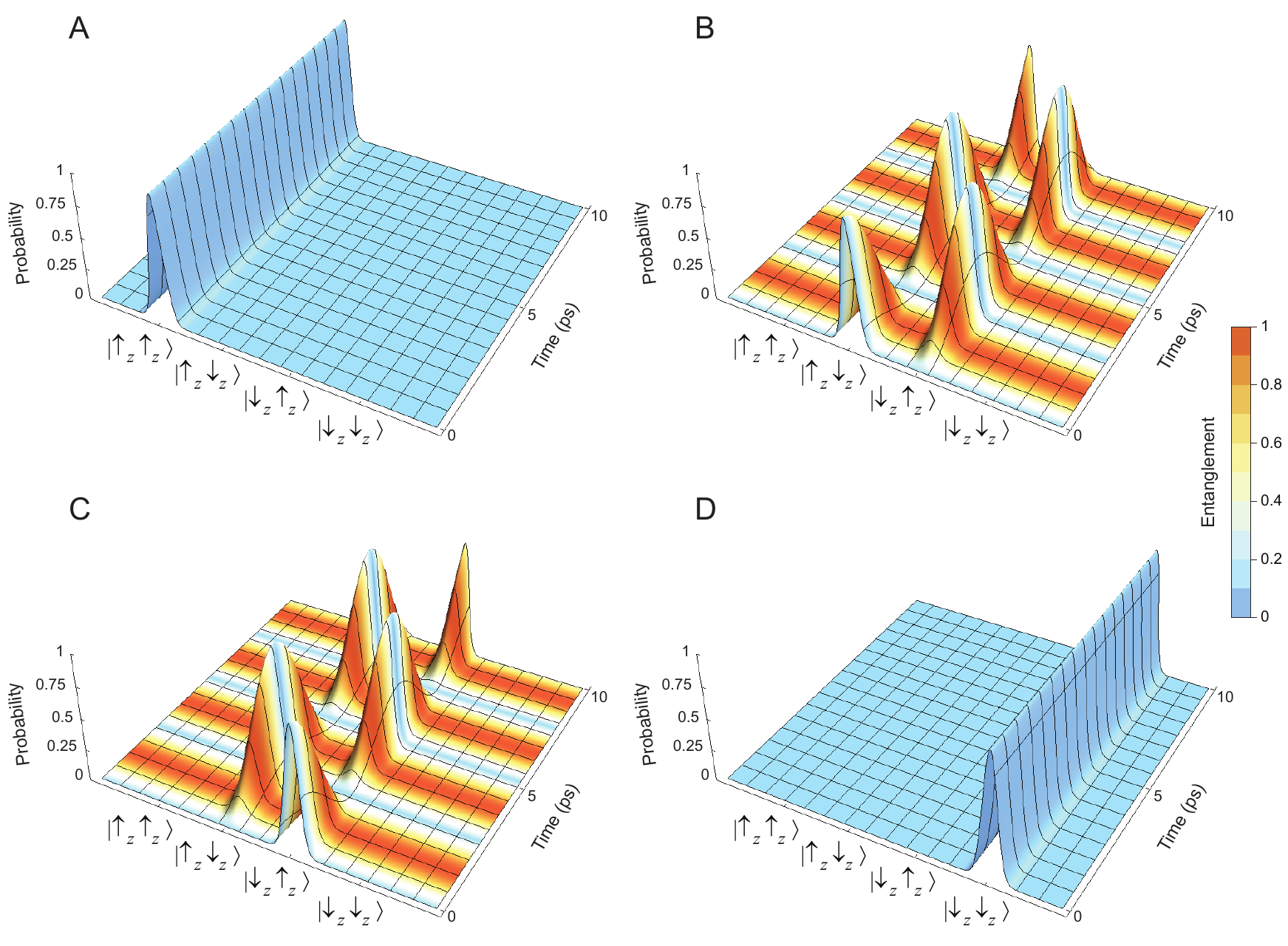}
\caption{\label{fig:2}Expectation values of the projectors
$\hat{\mathcal{P}}\left(\uparrow_{z}\uparrow_{z}\right)$,
$\hat{\mathcal{P}}\left(\uparrow_{z}\downarrow_{z}\right)$, $\hat{\mathcal{P}}\left(\downarrow_{z}\uparrow_{z}\right)$
and $\hat{\mathcal{P}}\left(\downarrow_{z}\downarrow_{z}\right)$
corresponding to probabilities of obtaining the given measurement
outcomes for the $z$-spin components of the two qubits. The~initial
state $|\Psi(0)\rangle$ at $t=0$ is $|\uparrow_{z}\uparrow_{z}\rangle$
in panel (\textbf{A}), $|\uparrow_{z}\downarrow_{z}\rangle$ in panel (\textbf{B}),
$|\downarrow_{z}\uparrow_{z}\rangle$ in panel (\textbf{C}) and $|\downarrow_{z}\downarrow_{z}\rangle$
in panel (\textbf{D}). The~internal Hamiltonians were modeled with \mbox{$\Omega_{1}=\Omega_{2}=0.3$
rad/ps}. The~interaction Hamiltonian was non-zero with \mbox{$\omega_{s}=0.3$~rad/ps}. The~amount of quantum entanglement at each moment of time
was measured using the normalized entanglement number $e(\Psi)/e(\Psi)_{\max}$.}
\end{figure}

In the presence of non-zero interaction Hamiltonian, the~quantum dynamics
of the composite state vector $|\Psi(t)\rangle$ is able to undergo
cycles of quantum entanglement and disentanglement depending on the
choice of the initial state vector $|\Psi(0)\rangle$. From \mbox{Equations~\eqref{eq:zz-1}--\eqref{eq:zz-4}}
describing the time evolution of the quantum probabilities, it can
be concluded that if the initial state is $|\Psi(0)\rangle=|\uparrow_{z}\uparrow_{z}\rangle$
then its associated observable represented by the expectation value
of the projection operator $\hat{\mathcal{P}}\left(\uparrow_{z}\uparrow_{z}\right)=|\uparrow_{z}\uparrow_{z}\rangle\langle\uparrow_{z}\uparrow_{z}|$
does not evolve but stays constant $\langle\Psi(t)|\hat{\mathcal{P}}\left(\uparrow_{z}\uparrow_{z}\right)|\Psi(t)\rangle=1$
at all times (Figure~\ref{fig:2}A). From \eqref{eq:sol-zz}, it can
be seen that the quantum evolution of the quantum state is to rotate
in the Hilbert space with a pure phase $e^{-\imath\omega_{1}t}|\uparrow_{z}\uparrow_{z}\rangle$,
which leaves the state separable at all times. It is worth emphasizing
that the quantum state evolves, whereas the expectation value of the
projection operator does not. This illustrates clearly the fact that
in the quantum world what is observed is not what exists, namely,
the quantum observables (represented by matrix operators) are not
quantum states (represented by state vectors). Similarly, if the initial
state is $|\Psi(0)\rangle=|\downarrow_{z}\downarrow_{z}\rangle$ then
its associated observable represented by the expectation value of
the projection operator $\hat{\mathcal{P}}\left(\downarrow_{z}\downarrow_{z}\right)=|\downarrow_{z}\downarrow_{z}\rangle\langle\downarrow_{z}\downarrow_{z}|$
does not evolve but stays constant $\langle\Psi(t)|\hat{\mathcal{P}}\left(\downarrow_{z}\downarrow_{z}\right)|\Psi(t)\rangle=1$
at all times. Again, the~quantum evolution of the quantum state is
to rotate in the Hilbert space with a pure phase $e^{-\imath\omega_{4}t}|\downarrow_{z}\downarrow_{z}\rangle$
thereby leaving the state separable at all times (Figure~\ref{fig:2}D).
Interesting quantum dynamics results when the initial state is $|\Psi(0)\rangle=|\uparrow_{z}\downarrow_{z}\rangle$
or $|\Psi(0)\rangle=|\downarrow_{z}\uparrow_{z}\rangle$. In such
cases, the~expectation values of the corresponding projection operators
evolve in time and there are observable quantum interference effects
that can be recorded by external measurement devices (Figure~\ref{fig:2}B,C).
Furthermore, the~state repeatedly gets entangled, when~$\textrm{Prob}\left(\uparrow_{z}\downarrow_{z}\right)=\textrm{Prob}\left(\downarrow_{z}\uparrow_{z}\right)=\frac{1}{2}$,
followed by disentanglement, when~one of the latter two probabilities
becomes unit and the other becomes zero. Thus, initially separable
composite quantum states are able to get quantum entangled if the
interaction Hamiltonian is non-zero.

\begin{figure}[t!]
\includegraphics[width=140mm]{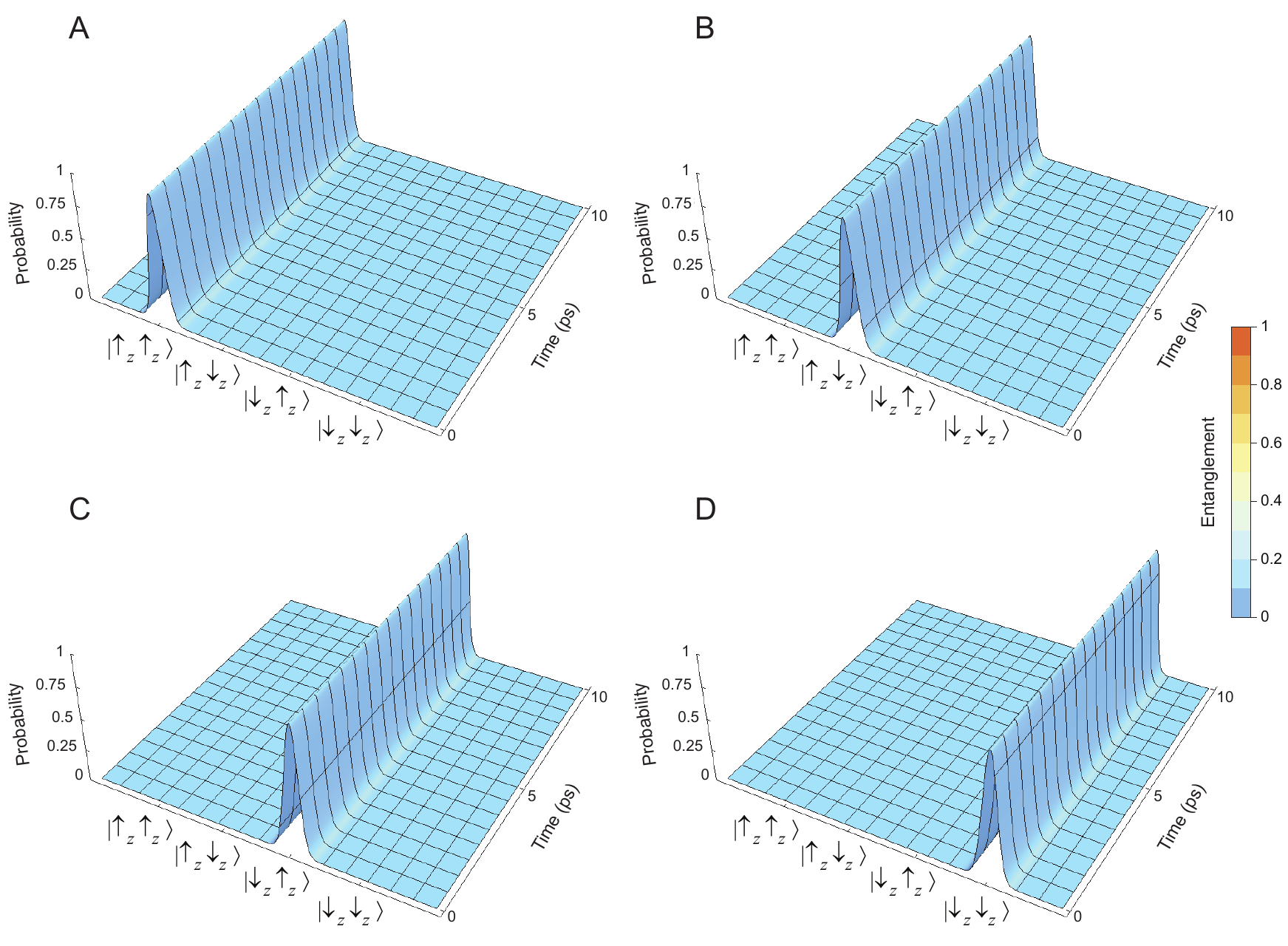}
\caption{\label{fig:3}Expectation values of the projectors
$\hat{\mathcal{P}}\left(\uparrow_{z}\uparrow_{z}\right)$,
$\hat{\mathcal{P}}\left(\uparrow_{z}\downarrow_{z}\right)$, $\hat{\mathcal{P}}\left(\downarrow_{z}\uparrow_{z}\right)$
and $\hat{\mathcal{P}}\left(\downarrow_{z}\downarrow_{z}\right)$
corresponding to probabilities of obtaining the given measurement
outcomes for the $z$-spin components of the two qubits. The~initial
state $|\Psi(0)\rangle$ at $t=0$ is $|\uparrow_{z}\uparrow_{z}\rangle$
in panel (\textbf{A}), $|\uparrow_{z}\downarrow_{z}\rangle$ in panel (\textbf{B}),
$|\downarrow_{z}\uparrow_{z}\rangle$ in panel (\textbf{C}) and $|\downarrow_{z}\downarrow_{z}\rangle$
in panel (\textbf{D}). The~internal Hamiltonians were modeled with $\Omega_{1}=\Omega_{2}=0.3$
rad/ps. The~interaction Hamiltonian was zero with $\omega_{s}=0$
rad/ps. The~amount of quantum entanglement at each moment of time
was measured using the normalized entanglement number $e(\Psi)/e(\Psi)_{\max}$.}
\end{figure}

In order to see what the role of the interaction Hamiltonian is, we
can turn it off by setting it to zero. In the absence of interaction
Hamiltonian, the~quantum dynamics of the composite state vector $|\Psi(t)\rangle$
is no longer able to undergo cycles of quantum entanglement and disentanglement
(Figure~\ref{fig:3}). Instead, the~initially separable quantum states
$|\Psi(0)\rangle$ remain separable at all times, as indicated by
the zero normalized entanglement number, $e(\Psi)/e(\Psi)_{\max}=0$.
Furthermore, the~expectation values stay constant 1 for the particular
projector $\hat{\mathcal{P}}\left(\uparrow_{z}\uparrow_{z}\right)$,
$\hat{\mathcal{P}}\left(\uparrow_{z}\downarrow_{z}\right)$, $\hat{\mathcal{P}}\left(\downarrow_{z}\uparrow_{z}\right)$
or $\hat{\mathcal{P}}\left(\downarrow_{z}\downarrow_{z}\right)$ that
corresponds to the initial state $|\Psi(0)\rangle$ (Figure~\ref{fig:3}A--D).
Thus, the~presence of non-zero interaction Hamiltonian is essential
for the generation of quantum entanglement starting from an initially
separable quantum state~$|\Psi(0)\rangle$.

By pairwise comparison of the computer plots performed with or without
non-zero interaction Hamiltonian, it can be seen that there is a difference
in the quantum dynamics if the initial state is $|\uparrow_{z}\downarrow_{z}\rangle$
as shown in \mbox{Figures \ref{fig:2}B and \ref{fig:3}B}, or $|\downarrow_{z}\uparrow_{z}\rangle$
as shown in \mbox{Figures \ref{fig:2}C and \ref{fig:3}C}. Yet, there is no
difference in the quantum dynamics if the initial state is $|\uparrow_{z}\uparrow_{z}\rangle$
as shown Figures \ref{fig:2}A and \ref{fig:3}A, or $|\downarrow_{z}\downarrow_{z}\rangle$
as shown in Figures \ref{fig:2}D and \ref{fig:3}D. The~explanation
for these findings is that the quantum dynamics for spin $zz$ states
is governed by the presence of non-zero off-diagonal elements in the
interaction Hamiltonian in the $zz$ basis
\begin{equation}
\hat{H}_{\textrm{int}}=\frac{\hbar^{2}}{4}\xi\left(\begin{array}{cccc}
1 & 0 & 0 & 0\\
0 & -1 & 2 & 0\\
0 & 2 & -1 & 0\\
0 & 0 & 0 & 1
\end{array}\right) . \label{eq:H-int}
\end{equation}
The first and fourth rows, which correspond respectively to the states
$|\uparrow_{z}\uparrow_{z}\rangle$ and $|\downarrow_{z}\downarrow_{z}\rangle$,
already contain off-diagonal zeroes, hence the zeroing of the interaction
Hamiltonian does not change anything. On the other hand, only the
second and third rows, which correspond respectively to the states
$|\uparrow_{z}\downarrow_{z}\rangle$ and $|\downarrow_{z}\uparrow_{z}\rangle$
contain non-zero off-diagonal elements (coherences). Zeroing of the
interaction Hamiltonian removes those non-zero off-diagonal elements,
which in turn prevents the two qubits from interacting with each other
and getting quantum entangled.

\subsection{Complementary Observables in Spin $xx$ Basis}

The non-commutativity of quantum operators results in incompatibility
of the corresponding observables, which cannot be measured with a
single experimental setting. Instead, measurements of incompatible
quantum observables require different mutually incompatible settings
of the employed experimental apparatus.

Suppose that we want to measure the two spins in $x$-direction. We
need to consider the relations
\begin{align}
\vert\uparrow_{z}\rangle & =\frac{1}{\sqrt{2}}\left(\vert\uparrow_{x}\rangle+\vert\downarrow_{x}\rangle\right)\\
\vert\downarrow_{z}\rangle & =\frac{1}{\sqrt{2}}\left(\vert\uparrow_{x}\rangle-\vert\downarrow_{x}\rangle\right).
\end{align}
Multiplying out the tensor products
\begin{align}
|\uparrow_{z}\uparrow_{z}\rangle & =\frac{1}{2}\left(\vert\uparrow_{x}\rangle+\vert\downarrow_{x}\rangle\right)\left(\vert\uparrow_{x}\rangle+\vert\downarrow_{x}\rangle\right)\\
|\uparrow_{z}\downarrow_{z}\rangle & =\frac{1}{2}\left(\vert\uparrow_{x}\rangle+\vert\downarrow_{x}\rangle\right)\left(\vert\uparrow_{x}\rangle-\vert\downarrow_{x}\rangle\right)\\
|\downarrow_{z}\uparrow_{z}\rangle & =\frac{1}{2}\left(\vert\uparrow_{x}\rangle-\vert\downarrow_{x}\rangle\right)\left(\vert\uparrow_{x}\rangle+\vert\downarrow_{x}\rangle\right)\\
|\downarrow_{z}\downarrow_{z}\rangle & =\frac{1}{2}\left(\vert\uparrow_{x}\rangle-\vert\downarrow_{x}\rangle\right)\left(\vert\uparrow_{x}\rangle-\vert\downarrow_{x}\rangle\right),
\end{align}
we obtain
\begin{align}
|\uparrow_{z}\uparrow_{z}\rangle & =\frac{1}{2}\left(\vert\uparrow_{x}\uparrow_{x}\rangle+\vert\uparrow_{x}\downarrow_{x}\rangle+\vert\downarrow_{x}\uparrow_{x}\rangle+\vert\downarrow_{x}\downarrow_{x}\rangle\right)\label{eq:zz-to-xx-1}\\
|\uparrow_{z}\downarrow_{z}\rangle & =\frac{1}{2}\left(\vert\uparrow_{x}\uparrow_{x}\rangle-\vert\uparrow_{x}\downarrow_{x}\rangle+\vert\downarrow_{x}\uparrow_{x}\rangle-\vert\downarrow_{x}\downarrow_{x}\rangle\right)\label{eq:zz-to-xx-2}\\
|\downarrow_{z}\uparrow_{z}\rangle & =\frac{1}{2}\left(\vert\uparrow_{x}\uparrow_{x}\rangle+\vert\uparrow_{x}\downarrow_{x}\rangle-\vert\downarrow_{x}\uparrow_{x}\rangle-\vert\downarrow_{x}\downarrow_{x}\rangle\right)\label{eq:zz-to-xx-3}\\
|\downarrow_{z}\downarrow_{z}\rangle & =\frac{1}{2}\left(\vert\uparrow_{x}\uparrow_{x}\rangle-\vert\uparrow_{x}\downarrow_{x}\rangle-\vert\downarrow_{x}\uparrow_{x}\rangle+\vert\downarrow_{x}\downarrow_{x}\rangle\right).\label{eq:zz-to-xx-4}
\end{align}
Upon setting $C_{\pm}=C_{2}\pm C_{1}$ and substitution of \eqref{eq:zz-to-xx-1}--\eqref{eq:zz-to-xx-4}
into \eqref{eq:sol-zz}, we obtain the quantum state in spin $xx$
basis
\begin{align}
|\Psi(t)\rangle & =\frac{1}{2}\left[\alpha_{1}(0)e^{-\imath\omega_{1}t}+C_{-}\alpha_{2}(0)e^{-\imath\omega_{2}t}+C_{+}\alpha_{3}(0)e^{-\imath\omega_{3}t}+\alpha_{4}(0)e^{-\imath\omega_{4}t}\right]\vert\uparrow_{x}\uparrow_{x}\rangle\nonumber \\
& \quad+\frac{1}{2}\left[\alpha_{1}(0)e^{-\imath\omega_{1}t}+C_{+}\alpha_{2}(0)e^{-\imath\omega_{2}t}-C_{-}\alpha_{3}(0)e^{-\imath\omega_{3}t}-\alpha_{4}(0)e^{-\imath\omega_{4}t}\right]\vert\uparrow_{x}\downarrow_{x}\rangle\nonumber \\
& \quad+\frac{1}{2}\left[\alpha_{1}(0)e^{-\imath\omega_{1}t}-C_{+}\alpha_{2}(0)e^{-\imath\omega_{2}t}+C_{-}\alpha_{3}(0)e^{-\imath\omega_{3}t}-\alpha_{4}(0)e^{-\imath\omega_{4}t}\right]\vert\downarrow_{x}\uparrow_{x}\rangle\nonumber \\
& \quad+\frac{1}{2}\left[\alpha_{1}(0)e^{-\imath\omega_{1}t}-C_{-}\alpha_{2}(0)e^{-\imath\omega_{2}t}-C_{+}\alpha_{3}(0)e^{-\imath\omega_{3}t}+\alpha_{4}(0)e^{-\imath\omega_{4}t}\right]\vert\downarrow_{x}\downarrow_{x}\rangle . \label{eq:sol-xx}
\end{align}
The probabilities of spin $xx$ observables are
\begin{align}
\textrm{Prob}\left(\uparrow_{x}\uparrow_{x}\right) & =\frac{1}{4}\left|\alpha_{1}(0)\right|^{2}+\frac{1}{2}C_{-}\left|\alpha_{1}(0)\alpha_{2}(0)\right|\cos\left[\left(\omega_{1}-\omega_{2}\right)t-\varphi_{1}+\varphi_{2}\right]\nonumber \\
& \quad+\frac{1}{2}C_{+}\left|\alpha_{1}(0)\alpha_{3}(0)\right|\cos\left[\left(\omega_{1}-\omega_{3}\right)t-\varphi_{1}+\varphi_{3}\right]\nonumber \\
& \quad+\frac{1}{2}\left|\alpha_{1}(0)\alpha_{4}(0)\right|\cos\left[\left(\omega_{1}-\omega_{4}\right)t-\varphi_{1}+\varphi_{4}\right]+\frac{1}{4}C_{-}^{2}\left|\alpha_{2}(0)\right|^{2}\nonumber \\
& \quad+\frac{1}{2}C_{+}C_{-}\left|\alpha_{2}(0)\alpha_{3}(0)\right|\cos\left[\left(\omega_{2}-\omega_{3}\right)t-\varphi_{2}+\varphi_{3}\right]\nonumber \\
& \quad+\frac{1}{2}C_{-}\left|\alpha_{2}(0)\alpha_{4}(0)\right|\cos\left[\left(\omega_{2}-\omega_{4}\right)t-\varphi_{2}+\varphi_{4}\right]+\frac{1}{4}C_{+}^{2}\left|\alpha_{3}(0)\right|^{2}\nonumber \\
& \quad+\frac{1}{2}C_{+}\left|\alpha_{3}(0)\alpha_{4}(0)\right|\cos\left[\left(\omega_{3}-\omega_{4}\right)t-\varphi_{3}+\varphi_{4}\right]+\frac{1}{4}\left|\alpha_{4}(0)\right|^{2}
\end{align}\vspace{-6pt}
\begin{align}
\textrm{Prob}\left(\uparrow_{x}\downarrow_{x}\right) & =\frac{1}{4}\left|\alpha_{1}(0)\right|^{2}+\frac{1}{2}C_{+}\left|\alpha_{1}(0)\alpha_{2}(0)\right|\cos\left[\left(\omega_{1}-\omega_{2}\right)t-\varphi_{1}+\varphi_{2}\right]\nonumber \\
& \quad-\frac{1}{2}C_{-}\left|\alpha_{1}(0)\alpha_{3}(0)\right|\cos\left[\left(\omega_{1}-\omega_{3}\right)t-\varphi_{1}+\varphi_{3}\right]\nonumber \\
& \quad-\frac{1}{2}\left|\alpha_{1}(0)\alpha_{4}(0)\right|\cos\left[\left(\omega_{1}-\omega_{4}\right)t-\varphi_{1}+\varphi_{4}\right]+\frac{1}{4}C_{+}^{2}\left|\alpha_{2}(0)\right|^{2}\nonumber \\
& \quad-\frac{1}{2}C_{+}C_{-}\left|\alpha_{2}(0)\alpha_{3}(0)\right|\cos\left[\left(\omega_{2}-\omega_{3}\right)t-\varphi_{2}+\varphi_{3}\right]\nonumber \\
& \quad-\frac{1}{2}C_{+}\left|\alpha_{2}(0)\alpha_{4}(0)\right|\cos\left[\left(\omega_{2}-\omega_{4}\right)t-\varphi_{2}+\varphi_{4}\right]+\frac{1}{4}C_{-}^{2}\left|\alpha_{3}(0)\right|^{2}\nonumber \\
& \quad+\frac{1}{2}C_{-}\left|\alpha_{3}(0)\alpha_{4}(0)\right|\cos\left[\left(\omega_{3}-\omega_{4}\right)t-\varphi_{3}+\varphi_{4}\right]+\frac{1}{4}\left|\alpha_{4}(0)\right|^{2}\vspace{-6pt}
\end{align}
\begin{align}
\textrm{Prob}\left(\downarrow_{x}\uparrow_{x}\right) & =\frac{1}{4}\left|\alpha_{1}(0)\right|^{2}-\frac{1}{2}C_{+}\left|\alpha_{1}(0)\alpha_{2}(0)\right|\cos\left[\left(\omega_{1}-\omega_{2}\right)t-\varphi_{1}+\varphi_{2}\right]\nonumber \\
& \quad+\frac{1}{2}C_{-}\left|\alpha_{1}(0)\alpha_{3}(0)\right|\cos\left[\left(\omega_{1}-\omega_{3}\right)t-\varphi_{1}+\varphi_{3}\right]\nonumber \\
& \quad-\frac{1}{2}\left|\alpha_{1}(0)\alpha_{4}(0)\right|\cos\left[\left(\omega_{1}-\omega_{4}\right)t-\varphi_{1}+\varphi_{4}\right]+\frac{1}{4}C_{+}^{2}\left|\alpha_{2}(0)\right|^{2}\nonumber \\
& \quad-\frac{1}{2}C_{+}C_{-}\left|\alpha_{2}(0)\alpha_{3}(0)\right|\cos\left[\left(\omega_{2}-\omega_{3}\right)t-\varphi_{2}+\varphi_{3}\right]\nonumber \\
& \quad+\frac{1}{2}C_{+}\left|\alpha_{2}(0)\alpha_{4}(0)\right|\cos\left[\left(\omega_{2}-\omega_{4}\right)t-\varphi_{2}+\varphi_{4}\right]+\frac{1}{4}C_{-}^{2}\left|\alpha_{3}(0)\right|^{2}\nonumber \\
& \quad-\frac{1}{2}C_{-}\left|\alpha_{3}(0)\alpha_{4}(0)\right|\cos\left[\left(\omega_{3}-\omega_{4}\right)t-\varphi_{3}+\varphi_{4}\right]+\frac{1}{4}\left|\alpha_{4}(0)\right|^{2}
\vspace{-6pt}\end{align}
\begin{align}
\textrm{Prob}\left(\downarrow_{x}\downarrow_{x}\right) & =\frac{1}{4}\left|\alpha_{1}(0)\right|^{2}-\frac{1}{2}C_{-}\left|\alpha_{1}(0)\alpha_{2}(0)\right|\cos\left[\left(\omega_{1}-\omega_{2}\right)t-\varphi_{1}+\varphi_{2}\right]\nonumber \\
& \quad-\frac{1}{2}C_{+}\left|\alpha_{1}(0)\alpha_{3}(0)\right|\cos\left[\left(\omega_{1}-\omega_{3}\right)t-\varphi_{1}+\varphi_{3}\right]\nonumber \\
& \quad+\frac{1}{2}\left|\alpha_{1}(0)\alpha_{4}(0)\right|\cos\left[\left(\omega_{1}-\omega_{4}\right)t-\varphi_{1}+\varphi_{4}\right]+\frac{1}{4}C_{-}^{2}\left|\alpha_{2}(0)\right|^{2}\nonumber \\
& \quad+\frac{1}{2}C_{+}C_{-}\left|\alpha_{2}(0)\alpha_{3}(0)\right|\cos\left[\left(\omega_{2}-\omega_{3}\right)t-\varphi_{2}+\varphi_{3}\right]\nonumber \\
& \quad-\frac{1}{2}C_{-}\left|\alpha_{2}(0)\alpha_{4}(0)\right|\cos\left[\left(\omega_{2}-\omega_{4}\right)t-\varphi_{2}+\varphi_{4}\right]+\frac{1}{4}C_{+}^{2}\left|\alpha_{3}(0)\right|^{2}\nonumber \\
& \quad-\frac{1}{2}C_{+}\left|\alpha_{3}(0)\alpha_{4}(0)\right|\cos\left[\left(\omega_{3}-\omega_{4}\right)t-\varphi_{3}+\varphi_{4}\right]+\frac{1}{4}\left|\alpha_{4}(0)\right|^{2}.
\end{align}

\begin{figure}[t!]
\includegraphics[width=140mm]{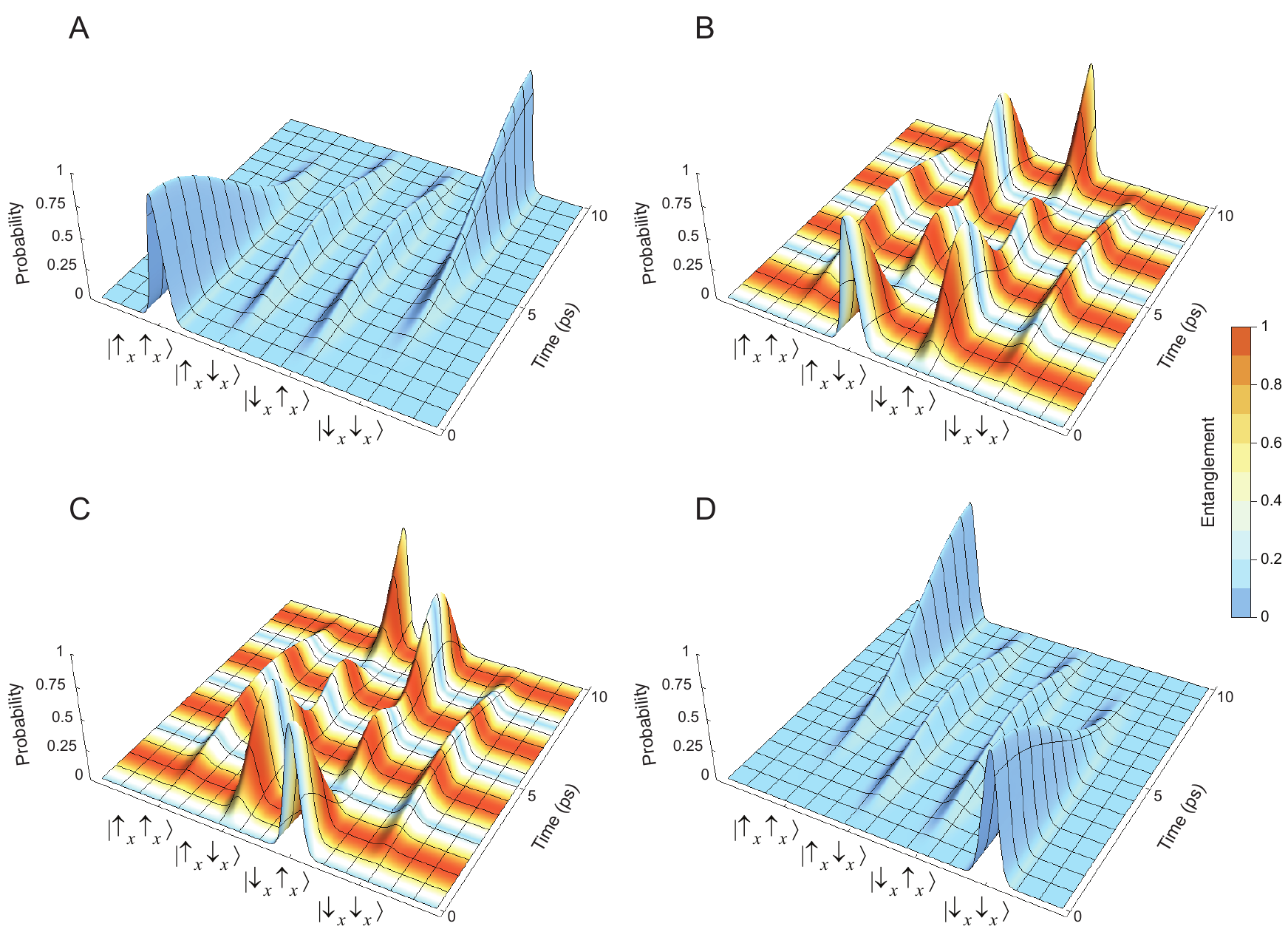}
\caption{\label{fig:4} Expectation values of the projectors
$\hat{\mathcal{P}}\left(\uparrow_{x}\uparrow_{x}\right)$,
$\hat{\mathcal{P}}\left(\uparrow_{x}\downarrow_{x}\right)$, $\hat{\mathcal{P}}\left(\downarrow_{x}\uparrow_{x}\right)$
and $\hat{\mathcal{P}}\left(\downarrow_{x}\downarrow_{x}\right)$
corresponding to probabilities of obtaining the given measurement
outcomes for the $x$-spin components of the two qubits. The~initial
state $|\Psi(0)\rangle$ at $t=0$ is $|\uparrow_{x}\uparrow_{x}\rangle$
in panel (\textbf{A}), $|\uparrow_{x}\downarrow_{x}\rangle$ in panel (\textbf{B}),
$|\downarrow_{x}\uparrow_{x}\rangle$ in panel (\textbf{C}) and $|\downarrow_{x}\downarrow_{x}\rangle$
in panel (\textbf{D}). The~internal Hamiltonians were modeled with \mbox{$\Omega_{1}=\Omega_{2}=0.3$~rad/ps}. The~interaction Hamiltonian was non-zero with $\omega_{s}=0.3$~rad/ps. The~amount of quantum entanglement at each moment of time
was measured using the normalized entanglement number $e(\Psi)/e(\Psi)_{\max}$.}
\end{figure}

Computer simulations confirm again that in the presence of non-zero
interaction Hamiltonian the quantum dynamics of the composite state
vector $|\Psi(t)\rangle$ is able to undergo cycles of quantum entanglement
and disentanglement depending on the choice of the initial state vector
$|\Psi(0)\rangle$ (Figure~\ref{fig:4}). When, the~initial state is
$|\Psi(0)\rangle=\vert\uparrow_{x}\uparrow_{x}\rangle$ (Figure~\ref{fig:4}A)
or $|\Psi(0)\rangle=\vert\downarrow_{x}\downarrow_{x}\rangle$ (Figure
\ref{fig:4}D), the~quantum dynamics leaves the state $|\Psi(t)\rangle$
separable at all times, as indicated by the zero normalized entanglement
number, $e(\Psi)/e(\Psi)_{\max}=0$. However, in these cases each
qubit manifests its own quantum interference pattern, which is independent
on the interference pattern manifested by the other qubit. This, reveals
the importance of the internal Hamiltonians to generate quantum interference
patterns only within the reduced subspaces of the component subsystems.
When, the~initial state is $|\Psi(0)\rangle=\vert\uparrow_{x}\downarrow_{x}\rangle$
(Figure~\ref{fig:4}B) or $|\Psi(0)\rangle=\vert\downarrow_{x}\uparrow_{x}\rangle$
(Figure~\ref{fig:4}C), the~quantum dynamics of the state $|\Psi(t)\rangle$
undergoes cycles of quantum entanglement and disentanglement. In the
presence of quantum entanglement, the~quantum interference patterns
manifested by the two qubits become correlated, which can be appreciated
by comparison with the case when the interaction Hamiltonian is turned
off.

\begin{figure}[t!]
\includegraphics[width=140mm]{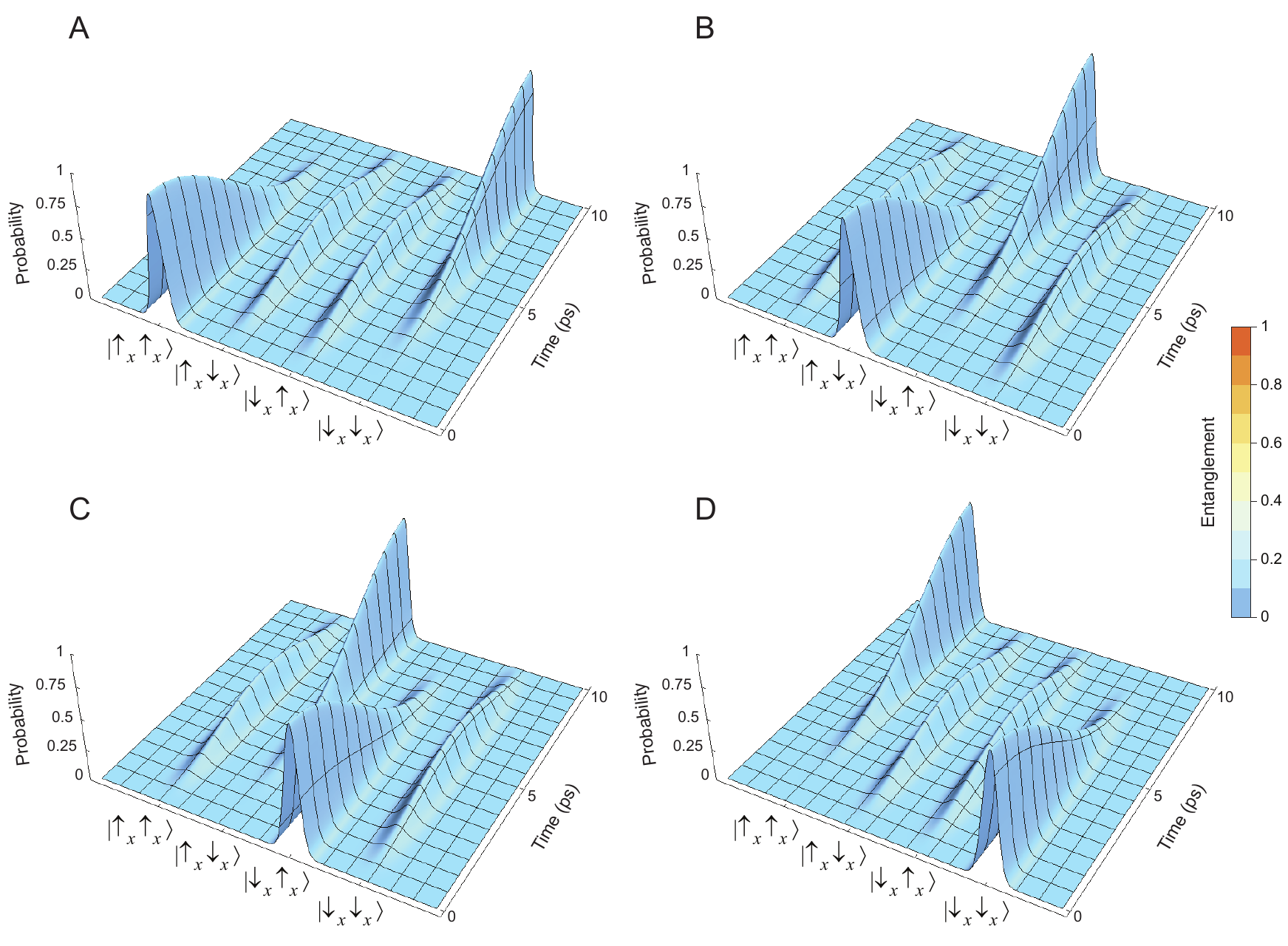}
\caption{\label{fig:5} Expectation values of the projectors
$\hat{\mathcal{P}}\left(\uparrow_{x}\uparrow_{x}\right)$,
$\hat{\mathcal{P}}\left(\uparrow_{x}\downarrow_{x}\right)$, $\hat{\mathcal{P}}\left(\downarrow_{x}\uparrow_{x}\right)$
and $\hat{\mathcal{P}}\left(\downarrow_{x}\downarrow_{x}\right)$
corresponding to probabilities of obtaining the given measurement
outcomes for the $x$-spin components of the two qubits. The~initial
state $|\Psi(0)\rangle$ at $t=0$ is $|\uparrow_{x}\uparrow_{x}\rangle$
in panel (\textbf{A}), $|\uparrow_{x}\downarrow_{x}\rangle$ in panel (\textbf{B}),
$|\downarrow_{x}\uparrow_{x}\rangle$ in panel (\textbf{C}) and $|\downarrow_{x}\downarrow_{x}\rangle$
in panel (\textbf{D}). The~internal Hamiltonians were modeled with $\Omega_{1}=\Omega_{2}=0.3$
rad/ps. The~interaction Hamiltonian was zero with $\omega_{s}=0$
rad/ps. The~amount of quantum entanglement at each moment of time
was measured using the normalized entanglement number $e(\Psi)/e(\Psi)_{\max}$.}
\end{figure}

In the absence of interaction Hamiltonian (Figure~\ref{fig:5}), the
quantum dynamics can no longer generate quantum entanglement from
initially separable composite quantum state $|\Psi(0)\rangle$. For
all four initial quantum states, the~quantum dynamics leaves the state
$|\Psi(t)\rangle$ separable at all times. When the initial state
is $|\Psi(0)\rangle=\vert\uparrow_{x}\uparrow_{x}\rangle$ (Figure~\ref{fig:5}A)
or \mbox{$|\Psi(0)\rangle=\vert\downarrow_{x}\downarrow_{x}\rangle$} (Figure
\ref{fig:5}D), the~expectation values for the projectors $\hat{\mathcal{P}}\left(\uparrow_{x}\uparrow_{x}\right)$,
$\hat{\mathcal{P}}\left(\uparrow_{x}\downarrow_{x}\right)$, $\hat{\mathcal{P}}\left(\downarrow_{x}\uparrow_{x}\right)$
and $\hat{\mathcal{P}}\left(\downarrow_{x}\downarrow_{x}\right)$
remain the same as in the corresponding cases with non-zero interaction
Hamiltonian (Figure~\ref{fig:4}A,D). However, when the initial state
is $|\Psi(0)\rangle=\vert\uparrow_{x}\downarrow_{x}\rangle$ \mbox{(Figure
\ref{fig:5}B)} or $|\Psi(0)\rangle=\vert\downarrow_{x}\uparrow_{x}\rangle$
\mbox{(Figure~\ref{fig:5}C),} the~expectation values for the projectors $\hat{\mathcal{P}}\left(\uparrow_{x}\uparrow_{x}\right)$,
$\hat{\mathcal{P}}\left(\uparrow_{x}\downarrow_{x}\right)$, $\hat{\mathcal{P}}\left(\downarrow_{x}\uparrow_{x}\right)$
and $\hat{\mathcal{P}}\left(\downarrow_{x}\downarrow_{x}\right)$
differ from those shown in the corresponding cases with non-zero interaction
Hamiltonian (Figure~\ref{fig:4}B,C). The~explanation of these findings
is based on the presence or absence of non-zero off-diagonal elements
(coherences) in the interaction Hamiltonian when expressed in the
spin $xx$ basis, analogously to the explanation that was already
given for the spin $zz$ observables. In fact, the~matrix form~\eqref{eq:H-int}
of the interaction Hamiltonian is preserved exactly the same during
the conversion from spin $zz$ basis to spin $xx$ basis. Noteworthy,
the internal Hamiltonians $\hat{H}_{A}$ and $\hat{H}_{B}$ do change
their matrix form during the conversion from spin $zz$ basis to spin
$xx$ basis, which in turn explains why the expectation values of
spin $xx$ or spin $zz$ observables for initial states where both
spins start aligned in the same direction, either display quantum
interference patterns (Figure \ref{fig:5}A,D) or stay constant (Figure
\ref{fig:3}A,D), respectively.

\section{\label{sec:disentangle}Quantum Dynamics of Initial Quantum Entangled
States}

\begin{figure}[t!]
\includegraphics[width=140mm]{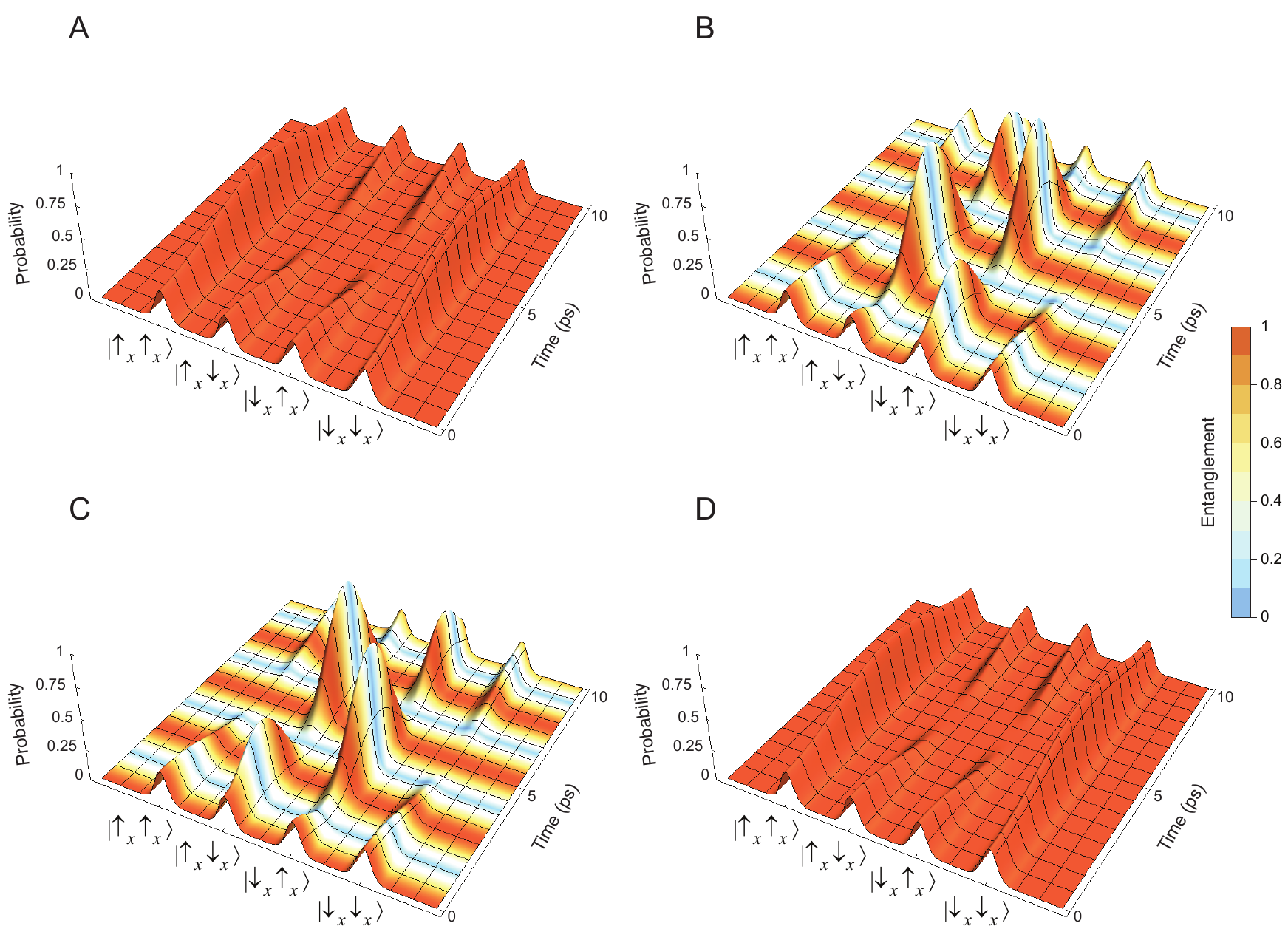}
\caption{\label{fig:6}Expectation values of the projectors $\hat{\mathcal{P}}\left(\uparrow_{x}\uparrow_{x}\right)$,
$\hat{\mathcal{P}}\left(\uparrow_{x}\downarrow_{x}\right)$, $\hat{\mathcal{P}}\left(\downarrow_{x}\uparrow_{x}\right)$
and $\hat{\mathcal{P}}\left(\downarrow_{x}\downarrow_{x}\right)$
corresponding to probabilities of obtaining the given measurement
outcomes for the $x$-spin components of the two qubits. The~initial
state $|\Psi(0)\rangle$ at $t=0$ is $\frac{1}{\sqrt{2}}\left(|\uparrow_{z}\uparrow_{x}\rangle+|\downarrow_{z}\downarrow_{x}\rangle\right)$
in panel (\textbf{A}), $\frac{1}{\sqrt{2}}\left(|\uparrow_{z}\downarrow_{x}\rangle+|\downarrow_{z}\uparrow_{x}\rangle\right)$
in panel (\textbf{B}), $\frac{1}{\sqrt{2}}\left(|\uparrow_{z}\uparrow_{x}\rangle-|\downarrow_{z}\downarrow_{x}\rangle\right)$
in panel (\textbf{C}) and $\frac{1}{\sqrt{2}}\left(|\uparrow_{z}\downarrow_{x}\rangle-|\downarrow_{z}\uparrow_{x}\rangle\right)$
in panel (\textbf{D}). The~internal Hamiltonians were modeled with $\Omega_{1}=\Omega_{2}=0.3$
rad/ps. The~interaction Hamiltonian was non-zero with $\omega_{s}=0.3$
rad/ps. The~amount of quantum entanglement at each moment of time
was measured using the normalized entanglement number $e(\Psi)/e(\Psi)_{\max}$.}
\end{figure}

The importance of non-zero interaction Hamiltonian for entangling
two qubits starting from an initially separable quantum state, goes
in the other direction as well. In other words, non-zero interaction
Hamiltonian is required to disentangle two qubits starting from an
initially entangled quantum state. For illustration, consider the
following four maximally entangled initial states $|\Psi(0)\rangle$
given by
\vspace{12pt}
\begin{align}
\frac{1}{\sqrt{2}}\left(|\uparrow_{z}\uparrow_{x}\rangle+|\downarrow_{z}\downarrow_{x}\rangle\right) & =\frac{1}{2}\left(|E_{1}\rangle+C_{-}|E_{2}\rangle+C_{+}|E_{3}\rangle-|E_{4}\rangle\right)\\
\frac{1}{\sqrt{2}}\left(|\uparrow_{z}\downarrow_{x}\rangle+|\downarrow_{z}\uparrow_{x}\rangle\right) & =\frac{1}{2}\left(|E_{1}\rangle+C_{+}|E_{2}\rangle-C_{-}|E_{3}\rangle+|E_{4}\rangle\right)\\
\frac{1}{\sqrt{2}}\left(|\uparrow_{z}\uparrow_{x}\rangle-|\downarrow_{z}\downarrow_{x}\rangle\right) & =\frac{1}{2}\left(|E_{1}\rangle-C_{+}|E_{2}\rangle+C_{-}|E_{3}\rangle+|E_{4}\rangle\right)\\
\frac{1}{\sqrt{2}}\left(|\uparrow_{z}\downarrow_{x}\rangle-|\downarrow_{z}\uparrow_{x}\rangle\right) & =\frac{1}{2}\left(|E_{1}\rangle-C_{-}|E_{2}\rangle-C_{+}|E_{3}\rangle-|E_{4}\rangle\right)
\end{align}
and suppose that we measure the expectation values of the projectors
$\hat{\mathcal{P}}\left(\uparrow_{x}\uparrow_{x}\right)$, $\hat{\mathcal{P}}\left(\uparrow_{x}\downarrow_{x}\right)$,
$\hat{\mathcal{P}}\left(\downarrow_{x}\uparrow_{x}\right)$ and $\hat{\mathcal{P}}\left(\downarrow_{x}\downarrow_{x}\right)$.
When the initial state $|\Psi(0)\rangle$ is $\frac{1}{\sqrt{2}}\left(|\uparrow_{z}\uparrow_{x}\rangle+|\downarrow_{z}\downarrow_{x}\rangle\right)$
(Figure~\ref{fig:6}A) or $\frac{1}{\sqrt{2}}\left(|\uparrow_{z}\downarrow_{x}\rangle-|\downarrow_{z}\uparrow_{x}\rangle\right)$
(Figure~\ref{fig:6}D), the~state $|\Psi(t)\rangle$ remains maximally
entangled at all times even though the spin $xx$ observables undergo
dynamics that manifests quantum interference patterns. However, when
the initial state $|\Psi(0)\rangle$ is $\frac{1}{\sqrt{2}}\left(|\uparrow_{z}\downarrow_{x}\rangle+|\downarrow_{z}\uparrow_{x}\rangle\right)$
\mbox{(Figure~\ref{fig:6}B)} or $\frac{1}{\sqrt{2}}\left(|\uparrow_{z}\uparrow_{x}\rangle-|\downarrow_{z}\downarrow_{x}\rangle\right)$
\mbox{(Figure~\ref{fig:6}C),} the~state $|\Psi(t)\rangle$ undergoes cycles
of disentanglement and entanglement. The~time points at which the
composite quantum state $|\Psi(t)\rangle$ becomes separable coincide
with unit quantum probability creating a peak at one of the two states
$|\uparrow_{x}\downarrow_{x}\rangle$ or $|\downarrow_{x}\uparrow_{x}\rangle$,
while all other orthogonal basis states remain empty with zero quantum~probability.

\begin{figure}[t!]
\includegraphics[width=140mm]{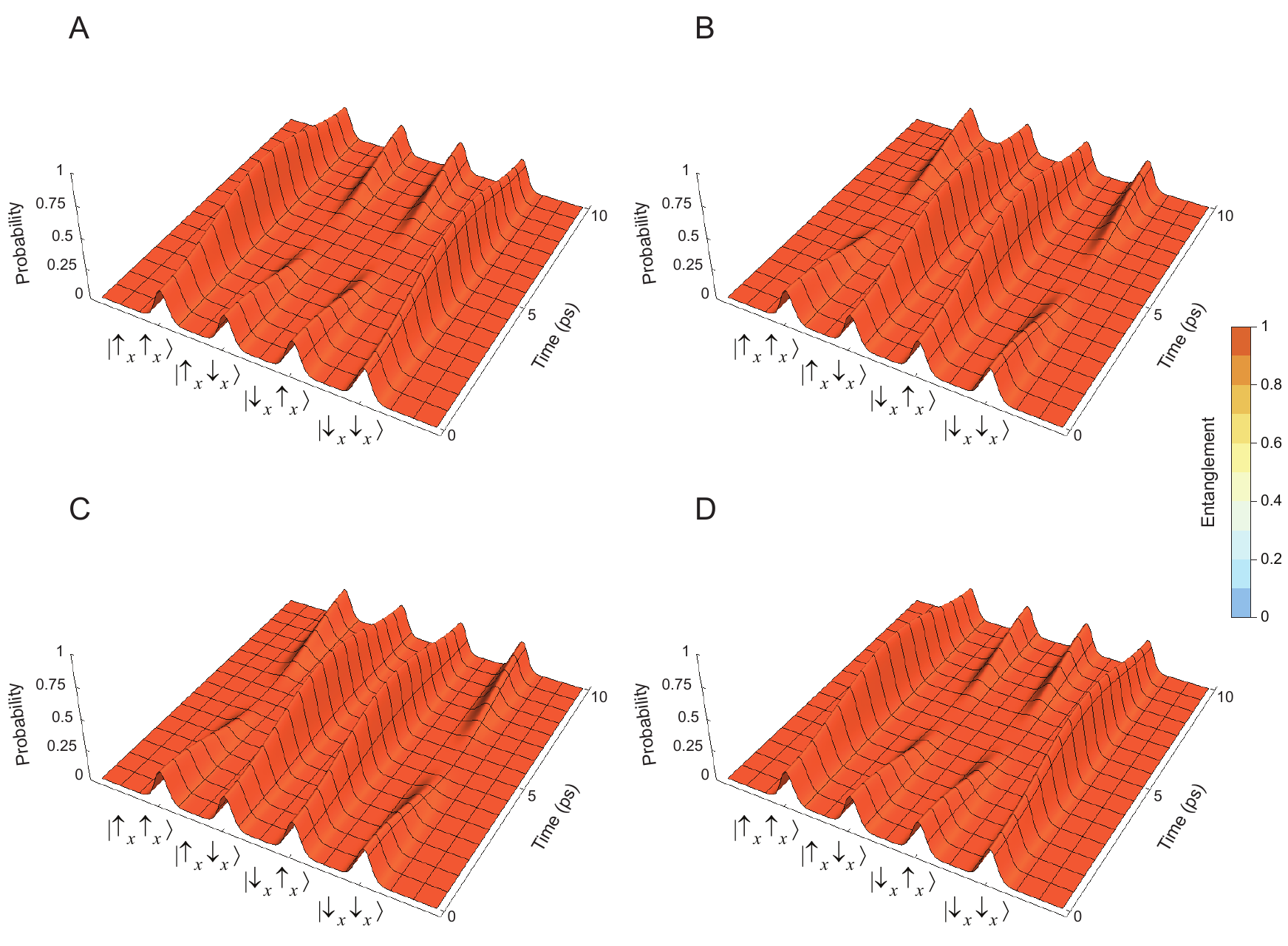}
\caption{\label{fig:7}Expectation values of the projectors $\hat{\mathcal{P}}\left(\uparrow_{x}\uparrow_{x}\right)$,
$\hat{\mathcal{P}}\left(\uparrow_{x}\downarrow_{x}\right)$, $\hat{\mathcal{P}}\left(\downarrow_{x}\uparrow_{x}\right)$
and $\hat{\mathcal{P}}\left(\downarrow_{x}\downarrow_{x}\right)$
corresponding to probabilities of obtaining the given measurement
outcomes for the $x$-spin components of the two qubits. The~initial
state $|\Psi(0)\rangle$ at $t=0$ is $\frac{1}{\sqrt{2}}\left(|\uparrow_{z}\uparrow_{x}\rangle+|\downarrow_{z}\downarrow_{x}\rangle\right)$
in panel (\textbf{A}), $\frac{1}{\sqrt{2}}\left(|\uparrow_{z}\downarrow_{x}\rangle+|\downarrow_{z}\uparrow_{x}\rangle\right)$
in panel (\textbf{B}), $\frac{1}{\sqrt{2}}\left(|\uparrow_{z}\uparrow_{x}\rangle-|\downarrow_{z}\downarrow_{x}\rangle\right)$
in panel (\textbf{C}) and $\frac{1}{\sqrt{2}}\left(|\uparrow_{z}\downarrow_{x}\rangle-|\downarrow_{z}\uparrow_{x}\rangle\right)$
in panel (\textbf{D}). The~internal Hamiltonians were modeled with $\Omega_{1}=\Omega_{2}=0.3$
rad/ps. The~interaction Hamiltonian was zero with $\omega_{s}=0$
rad/ps. The~amount of quantum entanglement at each moment of time
was measured using the normalized entanglement number $e(\Psi)/e(\Psi)_{\max}$.}
\end{figure}

\textls[-15]{In the case when the interaction Hamiltonian is turned off, all four
maximally entangled initial quantum states $\frac{1}{\sqrt{2}}\left(|\uparrow_{z}\uparrow_{x}\rangle+|\downarrow_{z}\downarrow_{x}\rangle\right)$
\mbox{(Figure~\ref{fig:7}A),} $\frac{1}{\sqrt{2}}\left(|\uparrow_{z}\downarrow_{x}\rangle+|\downarrow_{z}\uparrow_{x}\rangle\right)$
\mbox{(Figure~\ref{fig:7}B),} $\frac{1}{\sqrt{2}}\left(|\uparrow_{z}\uparrow_{x}\rangle-|\downarrow_{z}\downarrow_{x}\rangle\right)$
(Figure~\ref{fig:7}C) and $\frac{1}{\sqrt{2}}\left(|\uparrow_{z}\downarrow_{x}\rangle-|\downarrow_{z}\uparrow_{x}\rangle\right)$
\textls[-15]{(Figure~\ref{fig:7}D)} undergo quantum dynamics that leaves the composite
state $|\Psi(t)\rangle$ maximally entangled at all times. Now,~we~are ready to prove a general theorem according to which quantum dynamics
resulting only from internal Hamiltonians is unable to change the
amount of quantum entanglement that is already possessed by the initial
quantum state.}

\begin{Theorem}
Unitary quantum dynamics resulting from the Schr\"{o}dinger equation that
is only due to internal Hamiltonians is unable to change the amount
of quantum entanglement that is already present in the initial composite
quantum state $\vert\Psi(0)\rangle$.\end{Theorem}
\begin{proof}
Without loss of generality, suppose that we have a composite system
of two qubits governed by the Hamiltonian
\begin{equation}
\hat{H}=\hat{H}_{A}\otimes\hat{I}_{B}+\hat{I}_{A}\otimes\hat{H}_{B}.
\end{equation}
Express the initial quantum state in Schmidt basis
\begin{equation}
\vert\Psi(0)\rangle=\lambda_{1}\vert\uparrow_{A}\rangle\vert\uparrow_{B}\rangle+\lambda_{2}\vert\downarrow_{A}\rangle\vert\downarrow_{B}\rangle ,
\end{equation}
where the basis vectors are orthogonal
\begin{equation}
\langle\downarrow_{A}\vert\uparrow_{A}\rangle=\langle\downarrow_{B}\vert\uparrow_{B}\rangle=0.
\end{equation}
Solving the Schr\"{o}dinger equation gives the composite quantum state
at any time point $t$ using the matrix exponential of the Hamiltonian
\begin{equation}
\vert\Psi(t)\rangle=e^{-\frac{\imath}{\hbar}\left(\hat{H}_{A}\otimes\hat{I}_{B}+\hat{I}_{A}\otimes\hat{H}_{B}\right)t}\vert\Psi(0) \rangle .
\end{equation}
Using the commutativity of $\hat{H}_{A}\otimes\hat{I}_{B}$ and $\hat{I}_{A}\otimes\hat{H}_{B}$,
namely, $\left[\hat{H}_{A}\otimes\hat{I}_{B},\hat{I}_{A}\otimes\hat{H}_{B}\right]=0$,
we~can apply a special case of the Baker--Campbell--Hausdorff formula
\cite{Campbell1897,Baker1905,Strichartz1987,Nishimura2013}
\begin{equation}
e^{-\frac{\imath}{\hbar}\left(\hat{H}_{A}\otimes\hat{I}_{B}+\hat{I}_{A}\otimes\hat{H}_{B}+\frac{1}{2}\left[\hat{H}_{A}\otimes\hat{I}_{B},\hat{I}_{A}\otimes\hat{H}_{B}\right]\right)t}=e^{-\frac{\imath}{\hbar}\left(\hat{H}_{A}\otimes\hat{I}_{B}\right)t}e^{-\frac{\imath}{\hbar}\left(\hat{I}_{A}\otimes\hat{H}_{B}\right)t} .
\end{equation}
Using the power series definition of the matrix exponential, we have
\begin{equation}
\vert\Psi(t)\rangle=\sum_{k=0}^{\infty}\frac{1}{k!}\left[-\frac{\imath}{\hbar}\left(\hat{H}_{A}\otimes\hat{I}_{B}\right)t\right]^{k}\sum_{n=0}^{\infty}\frac{1}{n!}\left[-\frac{\imath}{\hbar}\left(\hat{I}_{A}\otimes\hat{H}_{B}\right)t\right]^{n}\vert\Psi(0)\rangle .
\end{equation}
Taking into account that powers of the identity operator are also
identity, we obtain
\begin{equation}
\vert\Psi(t)\rangle=\left(e^{-\frac{\imath}{\hbar}\hat{H}_{A}t}\otimes\hat{I}_{B}\right)\left(\hat{I}_{A}\otimes e^{-\frac{\imath}{\hbar}\hat{H}_{B}t}\right)\vert\Psi(0)\rangle=e^{-\frac{\imath}{\hbar}\hat{H}_{A}t}\otimes e^{-\frac{\imath}{\hbar}\hat{H}_{B}t}\vert\Psi(0)\rangle .
\end{equation}
Further substitution of the initial quantum state in Schmidt basis
followed by distribution of the time evolution operators gives
\begin{equation}
\vert\Psi(t)\rangle=\lambda_{1}e^{-\frac{\imath}{\hbar}\hat{H}_{A}t}\vert\uparrow_{A}\rangle\otimes e^{-\frac{\imath}{\hbar}\hat{H}_{B}t}\vert\uparrow_{B}\rangle+\lambda_{2}e^{-\frac{\imath}{\hbar}\hat{H}_{A}t}\vert\downarrow_{A}\rangle\otimes e^{-\frac{\imath}{\hbar}\hat{H}_{B}t}\vert\downarrow_{B}\rangle .
\end{equation}
Because both internal time evolution operators $e^{-\frac{\imath}{\hbar}\hat{H}_{A}t}$
and $e^{-\frac{\imath}{\hbar}\hat{H}_{B}t}$ are unitary, they preserve
inner products
\begin{align}
\langle\downarrow_{A}(t)\vert\uparrow_{A}(t)\rangle & =\langle\downarrow_{A}\vert e^{\frac{\imath}{\hbar}\hat{H}_{A}t}e^{-\frac{\imath}{\hbar}\hat{H}_{A}t}\vert\uparrow_{A}\rangle=\langle\downarrow_{A}\vert1\vert\uparrow_{A}\rangle=0\\
\langle\downarrow_{B}(t)\vert\uparrow_{B}(t)\rangle & =\langle\downarrow_{B}\vert e^{\frac{\imath}{\hbar}\hat{H}_{B}t}e^{-\frac{\imath}{\hbar}\hat{H}_{B}t}\vert\uparrow_{B}\rangle=\langle\downarrow_{B}\vert1\vert\uparrow_{B}\rangle=0.
\end{align}
This implies that the time evolved composite state decomposed in Schmidt
basis is
\begin{equation}
\vert\Psi(t)\rangle=\lambda_{1}\vert\uparrow_{A}(t)\rangle\vert\uparrow_{B}(t)\rangle+\lambda_{2}\vert\downarrow_{A}(t)\rangle\vert\downarrow_{B}(t)\rangle ,
\end{equation}
which has exactly the same Schmidt coefficients as $\vert\Psi(0)\rangle$.
The amount of quantum entanglement measured by the entanglement number
$e(\Psi)$ is only a function of the two Schmidt coefficients $\lambda_{1}$
and $\lambda_{2}$. Therefore, during quantum dynamics that is only
due to internal Hamiltonians, the~entanglement number stays constant,
$e\left[\Psi(0)\right]=e\left[\Psi(t)\right]$. Note~that we did not
resort anywhere in the proof to the fact that we have only two component
systems, or that the individual Hilbert spaces have only two complex
dimensions. Hence, the~presented proof straightforwardly generalizes
to multipartite systems with an arbitrary number of dimensions of
the individual Hilbert spaces.
\end{proof}

\section{\label{sec:condensation}Quantum Coherence Cannot Bind Conscious
Experiences}

Having explored the precise meaning of various quantum information-theoretic
concepts, we are ready to apply quantum information theory to a specific
problem of consciousness, namely, the~problem of binding
of conscious experiences.
\begin{Definition}
(Single mind) Awake healthy humans experience an integrated mental
picture consisting of visual, auditory, gustatory, olfactory, and
sensorimotor perceptions. Being a single mind can be understood introspectively
as the total content of all conscious experiences that you have at
a single time instant. Single minds are conscious \cite{Georgiev2017}.
\end{Definition}

\begin{Definition}
(Collection of minds) Being within a collection of minds can be understood
through contemplation of what it is like to be a participant in a
conversation with another human. When~you talk with a friend, you
may only guess what it is like to be inside your friend's mind, but
you do not have direct access to your friend's conscious experiences.
Thus, you and your friend are two separate minds. Taken together,
you and your friend, form a collection of minds. The~collection of
conscious minds is not conscious because it does not possess a single
mind \cite{Georgiev2017}.
\end{Definition}
Some quantum mind advocates use the phrase \emph{quantum coherence}
as a magical buzzword that presumably explains the physical mechanism
that binds your conscious experience, e.g., your visual experiences
with your auditory experiences, into a single integrated conscious
mind that is you. Unfortunately, this cannot be the case due to the
intimate relationship between subsystem coherence and separability of the composite system
as discussed in Sections \ref{sec:entanglement} and \ref{sec:coherence}.
The latter important point could be better appreciated by considering
the following illuminating example.
\begin{Example}
(Alice and Bob have two separate minds) Take two people, Alice and
Bob, each of which has an individual conscious mind. Let Alice have
a mind defined by the pure (quantum coherent) state $|\Psi_{A}\rangle$
and let Bob have a mind defined by the pure (quantum coherent) state
$|\Psi_{B}\rangle$. The~quantum mechanical axioms for composition
then imply that the composite system is also in a pure (quantum coherent)
state $|\Psi_{AB}\rangle=|\Psi_{A}\rangle\otimes|\Psi_{B}\rangle$.
Obviously, if all pure (quantum coherent) states had a conscious mind,
then $|\Psi_{AB}\rangle$ would be paradoxically a global conscious
mind as well \cite{Georgiev2017}. In order to ban paradoxical occurrence
of minds within other larger minds, quantum theory needs to postulate
that factorizable non-entangled quantum states correspond to a collection
of minds like Alice and Bob \cite{Georgiev2020a,Georgiev2020e}.
\end{Example}
\begin{Theorem}
Quantum purity (colloquially referred to as ``quantum coherence'' in decoherence theory) is related to separability, non-interaction and lack of binding, rather than binding of conscious experiences.
\end{Theorem}
\begin{proof}
Arrive at contradiction (i.e., occurrence of minds within other minds) by assuming the opposite (i.e., that quantum purity binds conscious experiences).
In mathematical logic, arriving at contradiction implies that the assumed premise is false.
\end{proof}
Extremely cold temperatures in quantum
technologies attempt to reduce the quantum interactions in order to
ensure that the initial composite state of qubits is separable in
the computational basis, and stays unaltered when no computational
quantum gates are performed \cite{Yang2020}. In other words, in order
for human programmer to be able to extract any useful information
from the quantum computation, the~quantum computer should be isolated
from interaction with its physical environment. However, we as conscious
minds do not exist in a void isolated from interaction from the rest
of the universe. On~the contrary, we constantly receive sensory information
from the surrounding world. What~is more, we enjoy so greatly being
aware of the surrounding world that \emph{total sensory deprivation}
of healthy human subjects in a dark room is experienced as a mental
torture \cite{Roth1987,Rasmussen1989,Reyes2007,Raz2013}. Furthermore,
there is no external programmer who is supposed to extract useful
information by observing the quantum state of our brain. Both of these
points show that even the motivation behind quantum mind models insisting
on preserving quantum coherence through putative isolation from the
environment is faulty. Quantum dynamics of brain molecules occurs
at picosecond timescale and rapid quantum entanglement due to quantum
interactions could lead to extensive correlations between different quantum
brain observables. Because composite quantum entangled states are
much more complex than separable states, it is quite natural to expect
that quantum entanglement due to quantum interactions is directly
related to the complexity of conscious experiences, whereas quantum
separability due to non-interaction of individually quantum coherent
components is related to disbinding or splitting of conscious experiences
into a collection of simpler, elementary minds \cite{Georgiev2020e}.
\begin{Definition}
(Kolmogorov complexity) The complexity $K$ of description of a given
string of symbols $S$ is the minimal number of classical information bits needed to
describe the string $S$ in some fixed universal description language \cite{Kolmogorov1968,Shen2017}.\end{Definition}
\begin{Example}
(Condensation reduces complexity) Modeling consciousness in terms
of \emph{condensation} of multiple components in the same quantum
state $\vert\psi\rangle\otimes\vert\psi\rangle\otimes\ldots\otimes\vert\psi\rangle$
inside some kind of room temperature superconductor (as suggested
in \cite{Marshall1989}) goes against the desirable link between complexity
and richness of conscious experiences. Complexity of quantum states
requires high level of entanglement as opposed to separability \cite{Susskind2020}.
Consider a simple system of 3 qutrits whose Hilbert space $\mathcal{H}$
is $3\times3\times3=27$ dimensional. To specify a general quantum
entangled state of these 3~qutrits, we will need 27 different complex
quantum probability amplitudes
\begin{equation}
|\Psi\rangle=\left(\begin{array}{c}
a_{1}\\
a_{2}\\
\vdots\\
a_{27}
\end{array}\right).
\end{equation}
If we suppose that for the encoding of each $a_{i}$ (e.g., as a root
of some polynomial equation) it takes $K$ bits of information, for
a general quantum entangled state we will need $27K$ bits.

If the state is factorizable, however, in the form
\begin{equation}
|\Psi\rangle=\left(\begin{array}{c}
a_{1}\\
a_{2}\\
a_{3}
\end{array}\right)\otimes\left(\begin{array}{c}
a_{4}\\
a_{5}\\
a_{6}
\end{array}\right)\otimes\left(\begin{array}{c}
a_{7}\\
a_{8}\\
a_{9}
\end{array}\right),
\end{equation}
then we will need only $3+3+3=9$ different complex quantum probability
amplitudes to fully specify the separable tensor product state. Thus,
for a general separable quantum state of 3 qutrits we will need $9K$ bits.

Finally, if all of the 3 qutrits are condensed in the same state
\begin{equation}
|\Psi\rangle=\left(\begin{array}{c}
a_{1}\\
a_{2}\\
a_{3}
\end{array}\right)\otimes\left(\begin{array}{c}
a_{1}\\
a_{2}\\
a_{3}
\end{array}\right)\otimes\left(\begin{array}{c}
a_{1}\\
a_{2}\\
a_{3}
\end{array}\right),
\end{equation}
then we will need only $3$ different complex quantum probability
amplitudes to fully specify the condensed state. The~resulting complexity
is only $3K$ bits. The~problem with condensation is obvious, namely,
the state becomes less and less complex, therefore conscious experiences
will become poorer and poorer as the condensation progresses. In fact,
the previous criticism with regard to quantum coherence and quantum
separability also applies to condensation. In particular, quantum
systems that are in a tensor product state are independent on each
other and it makes no sense to claim that they have a single mind,
on the contrary, there is a good reason to conclude that such quantum
systems correspond to a collection of separate, independent minds.
\end{Example}

\section{Conclusions}

Application of quantum information theory to studying neurophysiological processes is able to provide novel insights that are hard to anticipate from classical principles \mbox{alone \cite{Georgiev2017}}. To amend the
current lack of concise but rigorous introductions to quantum physics,
we~commenced this work with a brief description of the Hilbert space
formalism of quantum mechanics and explained its physical meaning
using a minimal quantum toy model. Then,~employing the tools of quantum
mechanics, we have determined the appropriate quantum dynamical timescale
of cognitive processes and pinpointed the physical mechanism that
binds conscious experiences within a single mind.

Our first main result is that the dynamic timescale of quantum processes,
which supports consciousness, has to be constrained by the
Planck--Einstein relation between energy and frequency. Because the
typical energies driving biological processes are greater than the
energy of the thermal noise, the~resulting dynamics is fixed to picosecond
timescale or faster, which is in the realm of quantum chemistry \cite{Lowe2005}.
This requires rethinking of the role of classical neuronal electric
spiking at millisecond timescale \cite{Rieke1999}, not as generating
elementary conscious events, but as a form of short term memory comprised
of billions of almost identical picosecond conscious events \cite{Georgiev2017}.
Of course, the~electric spike eventually fades away as the neuron
repolarizes, which means that the ongoing stream of consciousness
will forget the experiences that have triggered the spike unless another
biological form of storing long term memories is used, such as changing
the electric excitability of the neuronal membrane \cite{Zhang2003},
strengthening of the activity of the stimulated synapses \cite{Citri2007}
or morphological reorganization of the axo-dendritic neuronal trees
\cite{Chklovskii2004}.

Our second main result is that the quantum interactions due to non-zero
interaction Hamiltonian in the Schr\"{o}dinger equation are responsible
for dynamic changes in the amount of quantum entanglement, which leads
to decoherence of the individual quantum subsystems along with their
interaction. We also proved a rigorous quantum theorem according to
which unitary quantum dynamics that is only due to internal Hamiltonians
is unable to change the amount of quantum entanglement already present
in the initial composite quantum state. This implies that turning
off the interaction Hamiltonian will indeed preserve quantum coherence
of individual component subsystems starting from an initially separable
quantum state, however, the~quantum probabilities for different physical
observables measured on one of the subsystems will remain independent
from those measured on the other subsystem, which in turn precludes
any role of quantum coherence in cognitive binding of conscious experiences.
In fact, decomposition of a composite quantum state into a tensor
product of pure (quantum coherent) states is an indicator of splitting
of different conscious experiences into a collection of individual
conscious minds, whereas inseparability of a composite quantum entangled
state is a possible indicator of binding of different conscious experiences
into a single integrated conscious mind \cite{Georgiev2017}.

Studying the Schr\"{o}dinger equation may appear intimidating, but it is a profitable investment that pays off
with a high interest. The~mathematical constraints that enter into
the composition of a physically admissible Hamiltonian, together with
the axioms that relate quantum states with the expectation values
of quantum observables, are sufficient for the derivation of general
quantum information-theoretic no-go theorems that hold for all physical
processes including those that support consciousness \cite{Georgiev2013,Georgiev2017}.
These quantum theorems then could serve as theoretical tools to differentiate
between plausible and implausible physical solutions of open problems
in cognitive neuroscience.

%%%%%%%%%%%%%%%%%%%%%%%%%%%%%%%%%%%%%%%%%%
\vspace{6pt}

%%%%%%%%%%%%%%%%%%%%%%%%%%%%%%%%%%%%%%%%%%
%% optional
%\supplementary{The following are available online at \linksupplementary{s1}, Figure S1: title, Table S1: title, Video S1: title.}


%%%%%%%%%%%%%%%%%%%%%%%%%%%%%%%%%%%%%%%%%%
%\authorcontributions{For research articles with several authors, a short paragraph specifying their individual contributions must be provided. The~following statements should be used ``Conceptualization, X.X. and Y.Y.; methodology, X.X.; software, X.X.; validation, X.X., Y.Y. and Z.Z.; formal analysis, X.X.; investigation, X.X.; resources, X.X.; data curation, X.X.; writing---original draft preparation, X.X.; writing---review and editing, X.X.; visualization, X.X.; supervision, X.X.; project administration, X.X.; funding acquisition, Y.Y. All authors have read and agreed to the published version of the manuscript.'', please turn to the  \href{http://img.mdpi.org/data/contributor-role-instruction.pdf}{CRediT taxonomy} for the term explanation. Authorship must be limited to those who have contributed substantially to the work~reported.}

\funding{This research received no external funding.}
\institutionalreview{Not applicable.%In this section, please add the Institutional Review Board Statement and approval number for studies involving humans or animals. Please note that the Editorial Office might ask you for further information. Please add “The study was conducted according to the guidelines of the Declaration of Helsinki, and approved by the Institutional Review Board (or Ethics Committee) of NAME OF INSTITUTE (protocol code XXX and date of approval).” OR “Ethical review and approval were waived for this study, due to REASON (please provide a detailed justification).” OR “Not applicable” for studies not involving humans or animals. You might also choose to ex-clude this statement if the study did not involve humans or animals.
}

\informedconsent{Not applicable.%Any research article describing a study involving humans should contain this statement. Please add “Informed consent was obtained from all subjects involved in the study.” OR “Patient con-sent was waived due to REASON (please provide a detailed justification).” OR “Not applicable” for studies not involving humans. You might also choose to exclude this statement if the study did not involve humans. Written informed consent for publication must be obtained from participating patients who can be identified (including by the patients themselves). Please state “Written informed consent has been obtained from the patient(s) to publish this paper” if applicable.
}

\dataavailability{The morphology file of digitally reconstructed pyramidal neuron (NMO\_09565) from layer 5 of rat motor cortex is publicly available from NeuroMorpho.Org inventory (\url{http://NeuroMorpho.Org}; accessed on 19 April 2021).%In this section, please provide details regarding where data supporting reported results can be found, including links to publicly archived datasets analyzed or generated during the study. Please refer to suggested Data Availability Statements in section “MDPI Research Data Policies” at \href{https://www.mdpi.com/ethics}{https://www.mdpi.com/ethics}. You might choose to exclude this statement if the study did not report any data.
}

\conflictsofinterest{The author declares no conflict of interest.}

%% Optional
%\sampleavailability{Samples of the compounds ... are available from the authors.}

%%%%%%%%%%%%%%%%%%%%%%%%%%%%%%%%%%%%%%%%%%
%% Only for journal Encyclopedia
%\entrylink{The Link to this entry published on the encyclopedia platform.}

%%%%%%%%%%%%%%%%%%%%%%%%%%%%%%%%%%%%%%%%%%
%% Optional
%\abbreviations{The following abbreviations are used in this manuscript:\\
%\noindent
%\begin{tabular}{@{}ll}
%MDPI & Multidisciplinary Digital Publishing Institute\\
%DOAJ & Directory of open access journals\\
%TLA & Three letter acronym\\
%LD & Linear dichroism
%\end{tabular}}

%%%%%%%%%%%%%%%%%%%%%%%%%%%%%%%%%%%%%%%%%%
\end{paracol}
\reftitle{References}

\end{document}